\documentclass[longauth]{aa}
\usepackage[normalem]{ulem}
\usepackage{subfig}
\usepackage{graphicx}
\usepackage{adjustbox}
\usepackage{hyperref}
\usepackage{txfonts}
\usepackage[dvipsnames]{xcolor}
\usepackage{longtable}
\usepackage{caption}
\usepackage{pdflscape}
\usepackage{booktabs}
\usepackage{enumerate}
\usepackage{multirow}
\usepackage{soul}

\newcommand{\msun}{M$_\odot$}
\newcommand{\msunyr}{M$_\odot$\,yr$^{-1}$}
\newcommand{\kms}{km\,s$^{-1}$}

\makeatletter
\@ifpackageloaded{lineno}{\nolinenumbers\let\linenumbers\relax}{}%
\@ifpackageloaded{linenoaa}{\nolinenumbers\let\linenumbers\relax}{}%
\makeatother

\begin{document}

\title{The Nearby Evolved Stars Survey III: First data release of JCMT CO-line observations} 

\author{
S. H. J. Wallström \inst{1, 2}
\and P. Scicluna \inst{3, 4, 5, 2}
\and S. Srinivasan \inst{6, 2}\thanks{E-mail:s.srinivasan@irya.unam.mx}
\and J. G. A. Wouterloot \inst{7}
\and I. McDonald \inst{8, 9}
\and L. Decock \inst{1}
\and M. Wijshoff \inst{1}
\and R. Chen \inst{10, 2}
\and D. Torres \inst{11}
\and L. Umans \inst{1}
\and B. Willebrords \inst{1}
\and F. Kemper \inst{12, 13, 14}
\and G. Rau \inst{15, 16, 17}
\and S. Feng \inst{18, 2}
\and M. Jeste \inst{19}
\and T. Kaminski \inst{20}
\and D. Li \inst{21}
\and F. C. Liu \inst{22}
\and A. Trejo-Cruz \inst{6, 2}
\and H. Chawner \inst{23}
\and S. Goldman \inst{24, 25}
\and H. MacIsaac \inst{26, 27}
\and J. Tang \inst{21}
\and S. T. Zeegers \inst{28}
\and T. Danilovich \inst{29, 1}
\and M. Matsuura \inst{30}
\and K. M. Menten \inst{19}
\and J. Th van Loon \inst{31}
\and J. Cami \inst{26, 27, 32}
\and C. J. R. Clark \inst{24}
\and T. E. Dharmawardena \inst{33, 34}
\and J. Greaves \inst{23}
\and Jinhua He \inst{35, 36, 37}
\and H. Imai \inst{38, 39}
\and O. C. Jones \inst{40}
\and H. Kim \inst{41}
\and J. P. Marshall \inst{2}
\and H. Shinnaga \inst{42, 38}
\and R. Wesson \inst{30, 43}
\and the NESS Collaboration \inst{44}
}
\institute{
Institute of Astronomy, KU Leuven, Celestijnenlaan 200D bus 2401, 3001 Leuven, Belgium
\and Institute of Astronomy and Astrophysics, Academia Sinica, 11F of Astronomy-Mathematics Building, No.1, Sec. 4, Roosevelt Rd., Taipei 106319, Taiwan
\and Centre for Astrophysics Research, Department of Physics, Astronomy and Mathematics, College Lane Campus, University of Hertfordshire, Hatfield AL10 9AB, UK
\and European Southern Observatory, Alonso de Cordova 3107, Santiago RM, Chile
\and Space Science Institute, 4750 Walnut Street, Suite 205, Boulder, CO 80301, USA
\and Instituto de Radioastronomía y Astrofísica, Universidad Nacional Autónoma de México. Antigua Carretera a Pátzcuaro \#8701, Ex-Hda. San José de la Huerta 58089. Morelia, Michoacán, México
\and East Asian Observatory (JCMT), 660 N. A'ohoku Place, Hilo, HI 96720, USA
\and JBCA, Department Physics and Astronomy, University of Manchester, Manchester M13 9PL, UK
\and School of Physical Sciences, The Open University, Walton Hall, Milton Keynes, MK7 6AA, UK
\and Department of Physics, Duke University, Durham, NC 27708, USA
\and Departamento de Física, Matemáticas y Materiales, Universidad Autónoma de Ciudad Juárez, Ciudad Juárez, Chihuahua, Mexico
\and Institute of Space Science (ICE), CSIC, Can Magrans, E-08193 Cerdanyola del Vallès, Barcelona, Spain
\and ICREA, Pg. Lluís Companys 23, E-08010 Barcelona, Spain
\and Institut d'Estudis Espacials de Catalunya (IEEC), E-08860 Castelldefels, Barcelona, Spain
\and Schmidt Sciences, New York, NY 10011, USA
\and National Science Foundation, 2415 Eisenhower Avenue, Alexandria, Virginia 22314, USA
\and NASA Goddard Space Flight Center, Exoplanets and Stellar Astrophysics Laboratory, Code 667, Greenbelt, MD 20771, USA AND US National Science Foundation, Alexandria, VA
\and Department of Astronomy, Xiamen University, Zengcuo'an West Road, Xiamen, 361005 People's Republic of China
\and Max-Planck-Institut für Radioastronomie, 53121 Bonn, Germany
\and Nicolaus Copernicus Astronomical Center, Polish Academy of Sciences, Rabiánska 8, 87-100, Toruń, Poland
\and Xinjiang Astronomical Observatory, Chinese Academy of Sciences, Urumqi, 830011, People's Republic of China
\and National Taiwan Normal University, Earth Sciences, 88 Section 4, Ting-Chou Road, Taipei 11677, Taiwan
\and School of Physics and Astronomy, Cardiff University, Queen's Buildings, The Parade, Cardiff CF24 3AA, UK
\and Space Telescope Science Institute, 3700 San Martin Drive, Baltimore, MD 21218, USA
\and SOFIA-USRA, NASA Ames Research Center, MS 232-12, Moffett Field, CA 94035, USA
\and Department of Physics and Astronomy, The University of Western Ontario, London, ON, N6A 3K7, Canada
\and Institute for Earth and Space Exploration, The University of Western Ontario, London, ON, N6A 3K7, Canada
\and European Space Agency, ESTEC/SRE-SA, Keplerlaan 1, 2201 AZ, Noordwijk, The Netherlands
\and School of Physics and Astronomy, Monash University, Clayton 3800, Victoria, Australia
\and Cardiff Hub for Astrophysics Research and Technology (CHART), School of Physics and Astronomy, Cardiff University, The Parade, Cardiff CF24 3AA, UK
\and Lennard-Jones Laboratories, Keele University, ST5 5BG, UK
\and SETI Institute, 189 Bernardo Avenue, Suite 100, Mountain View, CA 94043, USA
\and Center for Computational Astrophysics, Flatiron Institute, 162 5th Ave., New York, NY 10010, USA
\and Center for Cosmology and Particle Physics, Department of Physics, New York University, 726 Broadway, New York, NY 10003, USA
\and Yunnan Observatories, Chinese Academy of Sciences, 396 Yangfangwang, Guandu District, Kunming 650216, China
\and Chinese Academy of Sciences South America Center for Astronomy, National Astronomical Observatories, CAS, Beijing 100101, China
\and Departamento de Astronomía, Universidad de Chile, Casilla 36-D, Santiago, Chile
\and Amanogawa Galaxy Astronomy Research Center, Graduate School of Science and Engineering, Kagoshima University, 1-21-35 Korimoto, Kagoshima, 890-0065, Japan
\and Center for General Education, Institute for Comprehensive Education, Kagoshima University, 1-21-30 Korimoto, Kagoshima, 890-0065, Japan
\and UK Astronomy Technology Centre, Royal Observatory, Blackford Hill, Edinburgh EH9 3HJ, UK
\and Korea Astronomy and Space Science Institute (KASI) 776, Daedeokdae-ro, Yuseong-gu, Daejeon 34055, Republic of Korea
\and Department of Physics and Astronomy, Kagoshima University, 1-21-35 Korimoto, Kagoshima, Japan
\and Department of Physics and Astronomy, University College London, Gower Street, London WC1E 6BT, UK
\and \url{https://evolvedstars.space/members}
}

\date{Received / Accepted }

\abstract 
{Low- to intermediate-mass ($\sim$0.8$-$8~M$_\odot$) evolved stars contribute significantly to the chemical enrichment of the interstellar medium in the local Universe. It is therefore crucial to accurately measure the mass return in their final evolutionary stages. The Nearby Evolved Stars Survey (NESS) is a large multi-telescope project targeting a volume-limited sample of $\sim$850 stars within 3~kpc in order to derive the dust and gas return rates in the Solar Neighbourhood, and to constrain the physics underlying these processes. We present an initial analysis of the CO-line observations, including detection statistics, carbon isotopic ratios, initial mass-loss rates, and gas-to-dust ratios. We describe a new data reduction pipeline for homogeneity, which we use to analyse the available NESS CO data from the James Clerk Maxwell Telescope, measuring line parameters and calculating empirical gas mass-loss rates. We present the first release of the available data on 485 sources, one of the largest homogeneous samples of CO data to date. Comparison with a large combined literature sample finds that high mass-loss rate and especially carbon-rich sources are over-represented in literature, while NESS is probing significantly more sources at low mass-loss rates, detecting 59 sources in CO for the first time and providing useful upper limits on non-detections. CO line detection rates are 81\% for the CO (2--1) line and 75\% for CO (3--2). The majority (82\%) of detected lines conform to the expected soft parabola shape, while eleven sources show a double wind. Calculated mass-loss rates show power-law relations with both the dust-production rates and expansion velocities, up to a mass-loss rate saturation value $\sim 5 \times 10^{-6}$~\msunyr. Median gas-to-dust ratios of 250 and 680 are found for oxygen-rich and carbon-rich sources, respectively. Our analysis of CO observations in this first data release highlights the importance of our volume-limited approach in characterizing the local AGB population as a whole.}

\maketitle

%%%%%%%%%%%%%%%%% BODY OF PAPER %%%%%%%%%%%%%%%%%%

\section{Introduction}

Cool evolved stars contribute to the chemical enrichment of the interstellar medium (ISM) by synthesising new elements  which are then expelled in massive stellar winds driven by newly formed dust. Population models indicate that low- to intermediate-mass ($\sim$0.8$-$8~M$_\odot$) asymptotic giant branch (AGB) stars dominate this process today \citep{karakas_dawes_2014}, aided by their more massive ($\geq$8~M$_\odot$) and therefore less numerous red supergiant (RSG) cousins, which eventually explode as supernovae. 
Accordingly, AGB stars are of great interest and many studies have been carried out on various aspects of their winds, including \citet{loup_co_1993, schoier_models_2001, olofsson_mass_2002, gonzalez_delgado_thermal_2003, ramstedt_circumstellar_2009} and \citet{de_beck_probing_2010}, which are described in Appendix~\ref{sect:lit_sample}.
Nevertheless, it has been difficult to draw firm conclusions about many key aspects of the population of AGB stars, including the total mass returned to the ISM by these stars, the physical
processes driving the onset of mass loss, the fraction of the ejected mass that condenses into dust, and variations in the mass-loss rate (MLR) over time \citep[e.g.,][]{hofner_mass_2018}. 

Repeated third dredge-up events during the AGB phase bring carbon-enriched material to the surface of the star, increasing its C/O ratio over time, and changing the dominant chemistry of their winds. AGB stars are accordingly divided into three groups that likely form an evolutionary sequence. The majority are oxygen-rich at solar metallicity and typically have C/O ratios $\lesssim 0.9$; their winds consist mainly of oxygen-bearing molecules and silicate dust. Carbon-rich AGB stars are at the other extreme (C/O $>$ 1), producing mainly carbonaceous molecules and amorphous carbon dust. S-type AGB stars have C/O ratios close to unity \citep[$0.9 \lesssim$ C/O $< 1.1$, with the exact lower limit being temperature-dependent; e.g.][]{ScaloRoss1976,vanEcketal2017} and ZrO bands stronger than TiO in low-resolution spectra \citep{Keenan1954}.

As AGB stars are major dust producers, we can use mid-IR observations to reveal their dust-forming regions and study their mass loss.
While distance uncertainties have hampered such studies in the Galaxy, the AGB populations of the Magellanic Clouds are well-studied. 
However, due to their low metallicity, the dominant dust producers in these galaxies are evolved carbon-rich stars \citep[e.g.,][]{riebel_mass-loss_2012,boyer_dust_2012,srinivasan_evolved-star_2016}. This limits our ability to extrapolate to the Galactic population, where oxygen-rich AGB stars are more common.
Furthermore, using dust to estimate the total return of enriched material to the ISM requires assumptions of the expansion velocity, dust-to-gas ratio, dust-grain properties and dust emissivity, which may differ between both individual stars and stellar populations, and may be function of metallicity. We therefore also want direct observations of the gas, which forms the majority of the lost mass.

CO observations are a good tracer of AGB gas mass loss \citep[e.g.,][]{kemper_mass_2003,decin_variable_2007} -- being both ubiquitous and chemically stable in AGB circumstellar shells, models show that its abundance relative to the main constituent of the stellar wind, H$_{2}$, varies by less than a factor of 2 across a range of C/O ratios \citep{cherchneff_chemical_2006}. In addition, it is very robust to photodissociation while its rotational transitions are excited at low temperatures\citep[e.g.][]{mamon_photodissociation_1988}. Together, these ensure it traces the bulk of the gas in the outflow within the photodissociation radius, without significant changes from source to source.
CO also provides a way to measure the $^{12}$C/$^{13}$C ratio through its $^{13}$CO lines. This ratio is modified by the nucleosynthesis in AGB stars \citep[e.g.,][]{kobayashi_evolution_2011} and hence, on Galactic scales, it can be used to trace star-formation histories. 
Excepting the brightest Magellanic Cloud objects \citep{groenewegen_alma_2016,matsuura_mass-loss_2016}, measurements of CO can only be made for Galactic sources, and its 
rotational transitions can only be systematically determined for Solar Neighbourhood objects (within a few kpc). 
In-depth studies of individual nearby sources continue to drive great progress in our understanding of the final stages of stellar evolution \citep[e.g.,][]{agundez_growth_2017, bujarrabal_structure_2021, hoai_morpho-kinematics_2022}, and broader studies have been done on limited samples or certain types of AGB star (e.g. \citealp{danilovich_new_2015, massalkhi_abundance_2018, wallstrom_atomium_2024} and the six studies in the literature sample described in Appendix~\ref{sect:lit_sample}). 
However, no large \emph{unbiased} study of Galactic AGB stars yet exists. 
As we cannot observe all Galactic AGB stars, a sample delineated only by geometry is the best way to circumvent potential biases and accurately characterize the population.

The {\it Nearby Evolved Stars Survey} \citep[NESS,][]{scicluna_nearby_2022} is a large, multi-telescope observing programme targeting a volume-limited sample of $\sim$850 mass-losing AGB and RSG stars within 3\,kpc for CO $J$=(2--1) and (3--2) spectral line and sub-millimetre (submm) continuum observations. 
The NESS sources are shown in Figure~\ref{fig:dist-DPR} and have been divided into the five tiers listed below, based on distance and dust-production rate (DPR), which is calculated by matching photometry with models from the Grid of Red supergiant and AGB ModelS \citep[GRAMS;][]{sargent_mass-loss_2011, srinivasan_mass-loss_2011} and detailed in \citet{scicluna_nearby_2022}. Note that some Galactic plane sources are excluded in more distant tiers due to the potential confusion from interstellar lines, and are instead part of a sample that must be observed separately with interferometry to filter out interstellar lines. The tiers are defined as follows:
\begin{enumerate}[(i)]
    \item "very low" sources with no detectable DPR, distances $d$ < 250 pc and luminosities of $L > 1600$ L$_\odot$, as measured by \citet{mcdonald_fundamental_2012} and \citet{mcdonald_fundamental_2017}; 
    \item "low" sources with $\mathrm{DPR} < 10^{-10}$ M$_\odot$\,yr$^{-1}$ and $d < 300$ pc; 
    \item "intermediate" sources with $10^{-10} < \mathrm{DPR} < 3 \times 10^{-9}$ M$_\odot$\,yr$^{-1}$ and $d < 600$ pc, excluding the Galactic plane for $d > 400$ pc (i.e., only including sources with  Galactic latitude $|b| > 1.5$) ; 
    \item "high" sources with $3 \times 10^{-9} < \mathrm{DPR} < 10^{-7}$ M$_\odot$\,yr$^{-1}$ and $d < 1200$ pc, excluding the Galactic plane for $d > 800$ pc; and 
    \item "extreme" sources with $\mathrm{DPR} > 10^{-7}$ M$_\odot$\,yr$^{-1}$ and $d < 3000$ pc, excluding the Galactic plane for $d > 2000$ pc.
\end{enumerate} 

In this paper, we present the initial CO results from observations of the northern part of the NESS sample with the James Clerk Maxwell Telescope (JCMT). 
Although incomplete, this initial dataset of 485 sources represents the largest sample of CO lines for nearby AGB stars to date, and includes 59 sources with no prior CO observations. Our highly homogeneous observations extend to systematically lower mass-loss rates than have typically been explored in the literature \citep{McDonald2025_NessCat}.
Section~\ref{sect:obs} describes the observations, data reduction pipeline, and data analysis, Section~\ref{sect:results} gives some initial results on the detection rates, line profiles, $^{12}$CO/$^{13}$CO ratios, mass-loss rates, gas-to-dust ratios, and comparisons with literature data and models, and finally Section~\ref{sect:conclusions} contains the conclusions.

\section{Observations and analysis}\label{sect:obs}

The NESS project is based around a JCMT Large Program to observe the $\sim$500 sources in the northern parts of the sample, for which we placed a declination cut at $-30^\circ$ to avoid observing at airmass $>$ 2. To this sample, we add all archival data for NESS targets taken with RxA3 and HARP since the ACSIS correlator was installed (from 2006 onwards). 
We note that this includes archival data of 71 sources that have Dec.~$< -30^\circ$. Because these sources are so far south, they have also been observed with APEX (along with other sources to ensure full-sky coverage), and sources observed by both telescopes will be compared in a paper analysing the APEX observations (Jeste et al., in prep).

The JCMT data presented here observed CO (2--1) and CO (3--2), as well as their $^{13}$CO isotopologues, in staring mode. We used double beam switching with a uniform chop throw of $180^{\prime\prime}$ at a position angle of $90^{\circ}$ East of North. There is also a small mapping subsample, and continuum observations at 850 and 450 $\mu$m, which will not be discussed in this work. See \citet{scicluna_nearby_2022} for more details on the observations. 

The NESS observations at the JCMT began in June 2017, and this paper includes all heterodyne data taken up to February 2023 using the RxA3/RxA3m\footnote{ RxA3 was the 3rd A-band receiver at the JCMT. It was upgraded after a mixer change in January 2016 to RxA3m, and throughout the rest of this paper we refer to both simply as RxA3.} and HARP \citep{buckle_harp/acsis:_2009} receivers. Note that the RxA3 receiver was removed from the JCMT in June 2018. Further observations, including with the new N\=amakanui receivers \citep{zmuidzinas_commissioning_2020} are underway and will complement the findings presented here.

\subsection{Removed sources}
We have removed a small number of sources from the analysis which, while meeting the original NESS sample criteria, are deemed to not be cool evolved stars. These sources are: IRAS 05251-1244, IRAS 06491-0654, IRAS 17150-3224, IRAS 17328-3327, IRAS 18458-0213, IRAS 19327+3024, IRAS 20002+3322, and IRAS 21282+5050. Most are planetary nebulae and one, IRAS 06491-0654, is not an evolved star at all but rather a Herbig Ae/Be star. 
Further details on the removed sources and the updated NESS sample are presented in \citet{McDonald2025_NessCat}.

\subsection{Main-beam efficiencies}\label{sect:etamb}

In order to convert each observation from a $T^*_\mathrm{A}$ to $T_\mathrm{mb}$ temperature scale, we have derived values for the main-beam efficiency ($\eta_\mathrm{mb}$) for both RxA3 and HARP using the full datasets of planet observations (Mars, Jupiter, and Uranus) from the JCMT. This data is presented graphically on the JCMT webpages for HARP\footnote{\url{https://www.eaobservatory.org/jcmt/instrumentation/heterodyne/calibration/harp-planets/}} and RxA3\footnote{\url{https://www.eaobservatory.org/jcmt/instrumentation/heterodyne/calibration/rxa3-planets/}}. Note that we do not use observations taken during the day (09--19$^h$ local time = 19--05$^h$ UT) as they tend to have larger uncertainties and systematically lower values of $\eta_\mathrm{mb}$. There have also been periods of misalignment of the RxA3 receiver, for each of which we calculate a separate best-fit $\eta_\mathrm{mb}$ value. An observational uncertainty, $\sigma_\eta$, was only available for the HARP observations, so we use the mean $\sigma_\eta$ value for observations with all receivers: $\sigma_\eta = 0.103 \pm 0.005$.

As the individual values of $\eta_\mathrm{mb}$ can vary widely, down to almost zero, due to various factors (including pointing errors, the amount of atmospheric water vapor, and uncertainties in the planet models), it can be difficult to ascertain the best values for $\eta_\mathrm{mb}$. For consistency, we elected to model the $\eta_\mathrm{mb}$ distribution as a function of time ($t$) as a straight line with an exponential bias term acting to reduce the measured values. The fitted function is as follows:
\begin{equation}\label{eq:fit_etamb}
    \eta_\mathrm{mb} = m*t + b - B*W + \epsilon 
\end{equation}
where $m$ is the (shallow) slope of the line, $b$ is the intercept, $\epsilon$ is Gaussian random noise such that $\epsilon\thicksim\mathcal{N}\left(0, \sigma_\eta\right)$ (where $\sigma_\eta$ is the measured uncertainty on the $\eta_\mathrm{mb}$ values), $B$ is the bias term that is drawn from an exponential distribution whose scale parameter $\sigma_B = 1/\lambda$ we infer, and $W$ is a weighting of the bias term (between 0 and 1). The use of an exponential distribution for the bias term results in asymmetry; this accounts for the systematic effects in observations that can reduce the \emph{measured} main-beam efficiency, e.g. bad pointing, which may produce underestimates of the true value on any given day. The values of $m, b, \sigma_B$, and $W$ were optimised by choosing a reasonable range for each parameter and then, for each of 1000 combinations of parameter values, randomly drawing 1000 distributions and comparing them to the distribution of measured $\eta_\mathrm{mb}$ values using the two-sample Kolmogorov--Smirnov (KS) test. The set of parameters with the lowest KS test statistic, corresponding to the smallest absolute difference between the empirical distribution functions of the model and the measured $\eta_\mathrm{mb}$ values, were taken to be the best-fit parameters.
The uncertainty on each parameter was calculated by finding the range of parameters which fell within the median absolute deviation of the KS test statistics, although note for the misalignment periods there were not enough  data points for this analysis. The slope was found to be 0 with an uncertainty of $\sim$10$^{-5}$ for all sets of observations, so the intercept $b$ is taken to be the value of $\eta_\mathrm{mb}$. The best-fit values of $\eta_\mathrm{mb}$, $\sigma_B$, and $W$ are given in Table~\ref{tab:etamb}. 

The best-fit $\eta_\mathrm{mb}$ values are used to correct each observational spectrum before they are combined into one spectrum (per frequency range) for each source. For later error propagation, a time-dependent, average error for the $\eta_\mathrm{mb}$ correction is used on spectral parameters derived from RxA3 observations. This error factor is equivalent to a fractional error of 0.21 in $\eta_\mathrm{mb}$.

\begin{table}[ht]
    \centering
    \caption{Best-fit values of the main beam efficiency ($\eta_\mathrm{mb}$, equal to $b$ in Equation~\ref{eq:fit_etamb}), width of the bias function ($\sigma_B$), and weight of the bias function ($W$) for the HARP ($325 - 375$~GHz) and RxA3 ($212 - 274$~GHz) receivers, including separate values for periods of misalignment (m), the dates of which are given as footnotes. Also given is the number of observations for each receiver/period. Uncertainties are given in parentheses, where they could be calculated; the uncertainties on $\eta_\mathrm{mb}$ also include the measurement uncertainty ($y_\mathrm{err}$).}
    \label{tab:etamb}
    \begin{tabular}{lcccc}
    \hline \hline
    Receiver & $\eta_\mathrm{mb}$ & $\sigma_B$ & $W$ & N$_\mathrm{obs}$\\
    \hline
    HARP & 0.57 (0.11) & 0.35 (0.01) & 0.45 (0.05) & 551 \\
    RxA3 & 0.60 (0.11) & 0.30 (0.01) & 0.30 (0.01) & 124 \\
    RxA3 m1 $^{(a)}$ & 0.45 (0.1) & 0.3 & 0.3 & 3 \\ 
    RxA3 m2 $^{(b)}$ & 0.53 (0.1) & 0.3 & 0.3 & 16 \\
    RxA3m $^{(c)}$ & 0.55 (0.14) & 0.35 (0.04) & 0.35 (0.04) & 64 \\
    RxA3m m$^{(d)}$ & 0.53 (0.1) & 0.4 & 0.4 & 6 \\
    \hline
    \multicolumn{4}{l}{\footnotesize $^{(a)}$ 20120413 -- 20121201;  $^{(b)}$ 20140508 -- 20150605 } \\ 
    \multicolumn{4}{l}{\footnotesize $^{(c)}$ From 20160101;  $^{(d)}$ 20170407 -- 20170810 } \\ 
    \end{tabular}
\end{table}

\subsection{Data reduction}

In order to obtain a homogeneous dataset, we have created an automated pipeline to reduce this large quantity of data, and perform an initial analysis. 
The aim of the NESS pipeline\footnote{Available at \url{https://github.com/swallstrom/JCMTpipeline}} is to reduce all existing heterodyne JCMT data of a given source, output a FITS image (single pixel for RxA3 data and 16 pixels for HARP) and a spectrum extracted from the primary pixel, fit any lines in the spectrum, and output a table of measured values. It is written in Python and makes use of the Starlink software \citep{currie_starlink_2014}.

\noindent For a given source, the pipeline will:
\begin{enumerate}[(i)]
    \item Query the JCMT archive for all observations matching the right ascension (R.A.) and Dec.\ of the source, with a given instrument, molecule, line, and observing mode, discarding observations marked either as "failed" or taken during the day (when the atmosphere is less stable). 
    \item Convert from $T^*_{\rm A}$ to $T_{\rm mb}$ temperature scale, using new $\eta_{\rm mb}$ determinations for both HARP and RxA3 (see Section~\ref{sect:etamb}). Perform a side-band correction for RxA3m data\footnote{see \href{https://www.eaobservatory.org/jcmt/wp-content/uploads/sites/2/2019/05/RxA3m-SB-Notes-2018.pdf}{https://www.eaobservatory.org/jcmt/wp-content/uploads/sites/2/2019/05/RxA3m-SB-Notes-2018.pdf}}.
    \item Reduce all raw files together in a group reduction using ORAC-DR \citep{jenness_automated_2015}, binning the data to channel widths of 1, 2, and 4 \kms.
    \item Output the group reduced file as a FITS image and extract the spectrum (in case of HARP observations, from the primary pixel).
    \item Fit a soft parabola function \citep[e.g.][]{olofsson_study_1993,de_beck_probing_2010} to the CO line (which is assumed to be the only bright line between --100 and +100 \kms) using the Markov-Chain Monte Carlo (MCMC) implementation \texttt{EMCEE} \citep{foreman-mackey_emcee:_2013} to quantify the uncertainties on the line peak, central velocity, width, and line shape parameters. All three spectral resolutions are fitted simultaneously under equal weighting.
    \item Calculate the root-mean square (RMS) noise of each spectrum using two regions, retaining the lower of the two values{\bf : i)} the region between --400 and --150 \kms, which does not contain any other commonly detected lines or likely ISM contamination and hence is assumed to be line-free; {\bf ii)} the region between --200 and 200 \kms\ after removing the fitted CO line, as this is the part of the spectrum that is manually inspected for, e.g., ISM contamination.  
    \item Output a data table containing the source coordinates, total integration time, RMS noise at each velocity resolution, and the best-fit line parameters, including positive and negative uncertainties.
\end{enumerate}

\subsection{Data analysis}\label{sect:data_analysis}

Each spectrum and line fit is manually checked to see if the line is fit reasonably well. A second pass of the MCMC line fitter, given input parameters from other lines in the same source, is carried out on spectra with a poor line fit. In the absence of other lines, the initial guesses for the parameters are estimated visually.

For the remainder of this manuscript, we use the 1~\kms\ spectra for display purposes.  
For each well-fit spectrum, the radial velocity and velocity width of the line is taken from the soft parabola fit parameters. These are used to define the velocity range over which to integrate the line (using the fit of the corresponding CO line to define the integration region for each $^{13}$CO line). We note that this may slightly underestimate the width of some lines which are less well fit by a soft parabola function, and for lines with clear line wings we instead take the line edges to be where the line emission falls below 3 $\times$ RMS. 
The integrated line intensities are used to calculate a $^{12}$CO/$^{13}$CO ratio, and an empirical MLR using the \citet{ramstedt_reliability_2008} formula:
\begin{equation}\label{eq:ramstedt}
    \mathrm{MLR} = s_J\  (I_\mathrm{CO}\: \theta^2_\mathrm{b}\: D^2)^{a_J}\  v^{b_J}_\mathrm{e}\  f^{-c_J}_\mathrm{CO}
\end{equation}
where $I_\mathrm{CO}$ is the integrated line intensity in K~km\,s$^{-1}$, $\theta_\mathrm{b}$ is the telescope beam size in arcseconds, $D$ is the distance in pc, $v_\mathrm{e}$ is the expansion velocity in km\,s$^{-1}$, $f_\mathrm{CO}$ is the CO abundance relative to H$_2$. The $s_J$, $a_J$, $b_J$, and $c_J$ parameters, and their uncertainties, are given for the fits to different CO transitions in their Table~A.1.
In cases where no $^{12}$CO line is detected (peak temperature $<$ 3 $\times$ RMS) we calculate an upper limit on the MLR, assuming an expansion velocity of 10~km\,s$^{-1}$ unless its value is known from other lines in the same source.

The integrated line intensities include uncertainties calculated from the rms noise of the spectrum, the uncertainty on $\eta_\mathrm{mb}$, the uncertainty in the line width, added in quadrature. 
The distance uncertainties \citep[for a discussion of their determination and uncertainties, see][]{scicluna_nearby_2022} are also propagated. 
The expansion velocity is taken to be half the line width, with an uncertainty equal to the channel width. 
$f_\mathrm{CO}$ is assumed to be $1 \times 10^{-3}$ for carbon-rich stars and $2 \times 10^{-4}$ for oxygen-rich stars \citep{ramstedt_reliability_2008}. The chemical type of the stars is determined as oxygen-rich or carbon-rich from mid-infrared spectra \citep[see][]{scicluna_nearby_2022} as a first approximation, and assumed to be oxygen-rich if no clear determination can be made. This causes S stars to be grouped with O-rich stars since they do not have strong SiC features. From the literature sample (see Section~\ref{sect:comparison_lit} and Appendix~\ref{sect:lit_sample}), there are 221 sources with unambiguous chemical classifications. Only 3/221 sources had classifications that conflicted with the mid-infrared spectroscopic classes: IRAS~20077-0625, IRAS~20141-2128, and IRAS~23438+0312. The chemical classifications for these sources have been fixed (now classified as O-rich, C-rich, and C-rich respectively), and we are confident the rest of the sample has similarly low error rates. However, further inspection of the chemical types is left to a future paper. 

To display distributions of MLR and gas-to-dust ratios in later figures, we show both histograms and kernel density estimates (KDEs). These KDEs are based on a Gaussian kernel, whose bandwidth is calculated using cross-validation by systematically testing a range of possible bandwidths. This calculated bandwidth is then added in quadrature to the median uncertainty on the plotted parameter (MLR or gas-to-dust ratio), so the resulting KDE is representative of the full uncertainty in the distribution.  

\section{Results and discussion}\label{sect:results}

A total of 485 sources from the full NESS sample have been included in this analysis: 259 sources with CO (2--1) observations, and 428 sources with CO (3--2) observations.
The full analysis results, with measured parameters and calculated values, are available online at the CDS via Vizier. Table~\ref{tab:analysis_results} lists the columns available in the online dataset.

\subsection{Detection statistics}\label{sect:detection_stats}

\begin{figure}
    \centering
    \includegraphics[width=\columnwidth]{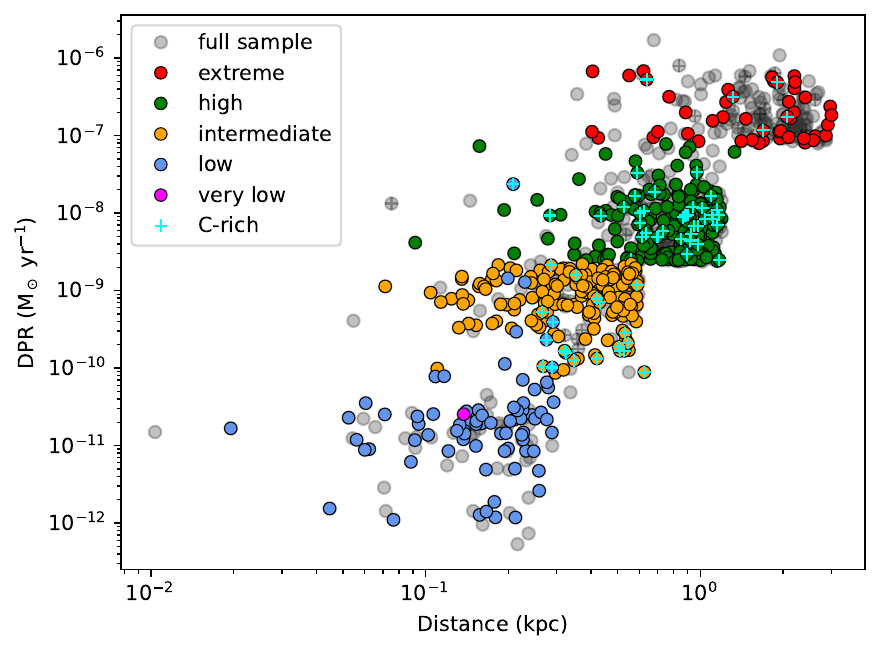}
    \caption{Distance vs.\ dust production rate (DPR) for the full NESS sample, in grey. Overlaid are the sources which have been observed and are included in the current analysis, coloured by tier. Cyan crosses show the locations of the carbon-rich sources.}
    \label{fig:dist-DPR}
\end{figure}

Figure~\ref{fig:dist-DPR} shows the full NESS sample and the sources that are included in the current analysis. The three middle tiers ("low", "intermediate", and "high") are well sampled ($\sim$60--75\% observed), while fewer sources in the "extreme" tier have been observed ($\sim$30\%) so far. A larger fraction of the less numerous and generally brighter carbon-rich sources in NESS have been observed so far, compared to oxygen-rich sources.

The observation and detection statistics for the NESS sample so far, both in total and divided by tier, are shown in Table~\ref{tab:detections} and Fig.~\ref{fig:detection-stats} respectively.
In total, we detect about 80\% of targeted sources in CO (2--1) and 75\% in CO (3--2). Of these, 59 sources appear to have no previously published CO observations; these sources are marked in the online table.
Detection statistics for $^{13}$CO are around 40\% for $^{13}$CO (2--1) and 30\% for $^{13}$CO (3--2). We note that $^{13}$CO has not been targeted towards all sources with the JCMT, as observing priority has so far been given to sources with $^{12}$CO (2--1) or (3--2) detections brighter than 0.3~K. 
The "low" tier has lower detection rates than the other tiers. This is unsurprising as it includes sources with no measurable DPR, which may not be able to launch a significant dust-driven wind at all. Detection statistics for the three highest tiers are comparable, despite relatively fewer sources having been observed so far in the most distant tiers ("high" and "extreme"), indicating that our observing strategy is not biased towards a particular type of source. 
We also note that, overall, the C-rich sources show higher detection rates, consistent with their generally higher MLRs.

\begin{table}
\caption{Fraction of the overall NESS sample that has been observed with the JCMT in this work, divided by tier and chemical type.}
\label{tab:detections} 
\centering 
\begin{tabular}{lrrr}
\hline\hline
Tier & Total & O-rich$^{(a)}$ & C-rich \\
\hline
All tiers & 485/852 (57\%) & 421/757 (56\%) & 64/95 (67\%) \\
Very low & 1/19 (5\%) & 1/19 (5\%) & 0/0 (--) \\
Low & 76/105 (72\%) & 72/101 (71\%) & 4/4 (100\%) \\
Interm. & 169/222 (76\%) & 152/201 (76\%) & 17/21 (81\%) \\
High & 190/324 (59\%) & 152/273 (56\%) & 38/51 (75\%) \\
Extreme & 49/182 (27\%) & 44/163 (27\%) & 5/19 (26\%) \\
\hline\hline
\end{tabular}
\parbox[t]{\columnwidth}{
\footnotesize $^{(a)}$ sources without mid-IR based chemical identifications are assumed\\ to be O-rich; see Section \ref{sect:data_analysis} for details.
}
\end{table}
\subsection{Line profiles}\label{sect:line_profiles}

% \begin{figure}
%     \centering
%     \includegraphics[width=\columnwidth]{detstats_heatmap_O-rich.pdf}\\
%     \includegraphics[width=\columnwidth]{detstats_heatmap_C-rich.pdf}\\
%     \includegraphics[width=\columnwidth]{detstats_heatmap_Total.pdf}
%     \caption{Heatmaps showing detections in the JCMT heterodyne data processed in this paper for the O--rich (top), C--rich (centre), and the full sample of observed NESS sources.}
%     \label{fig:detection-stats}
% \end{figure}

All $^{12}$CO spectra have been judged by eye on whether they follow a soft parabola shape, as is expected for a constant-velocity, spherically symmetric wind. Some $\sim$7\% of detected lines are too weak for their shape to be determined. However, of the sources with sufficiently bright CO lines, 82\% conform to a soft parabola shape. For O-rich sources, 85\% show a soft parabola shape, compared with a slightly smaller fraction, 74\%, of C-rich sources.
These proportions are essentially upper limits, as we have assumed a soft parabola shape as the default and there may be sources whose deviations from that shape are less clearly distinguishable due to noise.

\begin{figure*}
	\includegraphics[width=\textwidth]{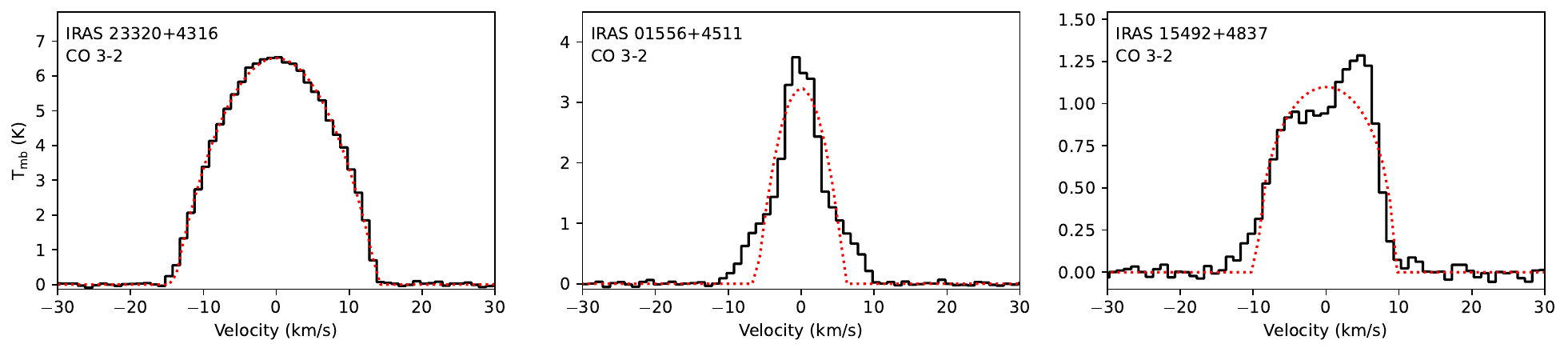}
    \caption{Three typical examples of the different CO line shapes, with the best-fit soft parabola shown with a red dotted line. From left to right: LP And (IRAS 23320+4316)  shows a soft parabola profile, V360 And (IRAS 01556+4511) shows a double-wind profile, and ST Her (IRAS 15492+4837) shows an asymmetric profile with line wings.}
    \label{fig:nonSP}
\end{figure*}

\begin{figure}
    \centering
    \includegraphics[width=\columnwidth]{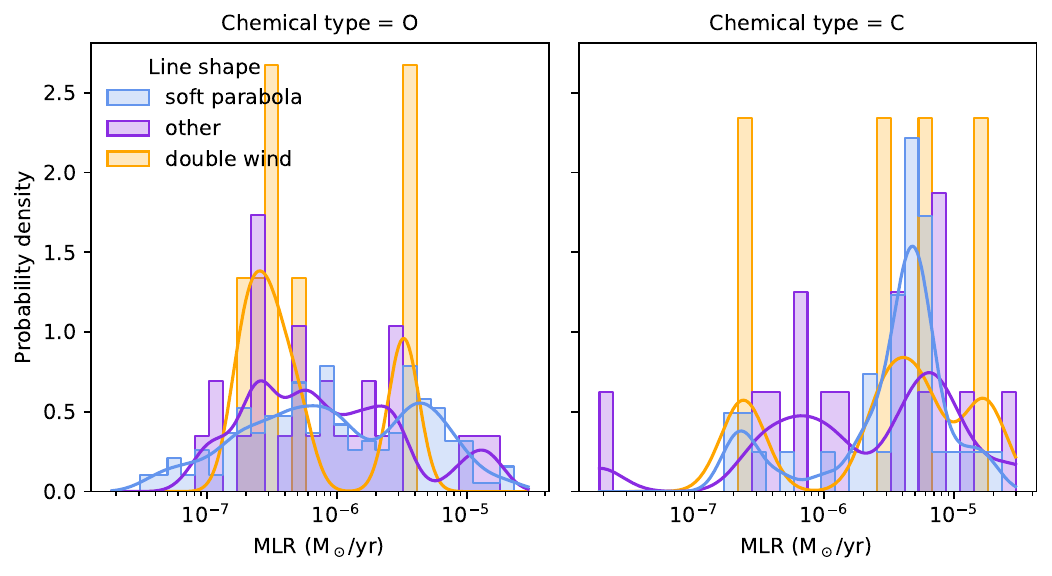}
    \caption{Distributions of MLRs derived from CO lines with a soft parabola shape, a double wind, or other non-soft-parabola shape.}
    \label{fig:sp_not}
\end{figure}

Three typical examples of the different CO line shapes are shown in Figure~\ref{fig:nonSP}, and a histogram and KDE of the MLRs of sources with different line shapes is shown in Figure~\ref{fig:sp_not}. The O-rich sources show a large spread in MLR, while the fewer C-rich sources are concentrated at higher MLRs, peaking around $5 \times 10^{-6}$~\msunyr.
Within each chemistry, the distributions of MLRs for different line profiles are comparable. There is a possible indication of lower typical MLRs for O-rich stars with non-soft parabola line shapes, but the number of such sources is too low to support a definitive conclusion.
The most common deviations from the soft parabola shape are asymmetry and/or line wings. Furthermore, 11 of the 485 sources clearly show multiple velocity components centred on the same velocity: double winds. 
The majority (7/11) of these double-wind sources are O-rich, and they cluster around a low MLR of a few $\times$ 10$^{-7}$~\msunyr. The C-rich double wind sources have a range of MLRs, between $2.2 \times 10^{-7}$ and $1.7 \times 10^{-5}$~\msunyr. 

Our results are similar to what was found by  \citet{knapp_multiple_1998} for a sample of 43 CO-bright AGB stars: 30 (70\%) of their sources show a parabolic line shape, six are asymmetric, and seven show double winds. Of their 13 C-rich sources, one is asymmetric and one shows an uncertain double wind, so 85\% show a parabolic line shape. They also find that their stars with double winds tend to have relatively low MLRs, largely due to the slower wind component, which also has a lower expansion velocity than most AGB winds, and speculate that this corresponds to a slow wind following a period of increased mass loss. 

\subsection{Carbon isotopic ratios}\label{sect:isorat}

The $^{12}$C/$^{13}$C ratio is a tracer of the evolutionary state and nucleosynthesis in AGB stars \citep[e.g.][]{milam_oxygen-rich_2007, ramstedt_12co13co_2014}. The previous evolution on the red giant branch tends to produce low $^{12}$C/$^{13}$C ratios, around 5$-$10 \citep{pavlenko_carbon_2003}, especially for low-mass stars. Then, during the AGB phase, dredge-ups will increase the amount of carbon (and specifically $^{12}$C) in the surface layers of the star, which also increases the $^{12}$C/$^{13}$C ratio. A further effect takes place in the more massive AGB stars, above $\sim$4~\msun, where hot-bottom burning will consume $^{12}$C and lower the $^{12}$C/$^{13}$C ratio \citep{karakas_dawes_2014}. 
From these evolutionary processes, we would na\"ively expect a correlation between $^{12}$C/$^{13}$C ratio and MLR. Based on the above processes and a reasonable IMF, we expect that most early AGB stars have low $^{12}$C/$^{13}$C ratios and low MLRs, and both quantities tend to increase during the AGB phase. Hot-bottom burning lowers the $^{12}$C/$^{13}$C ratios in the rarer intermediate-mass AGB stars, which generally have higher MLRs.

\begin{figure}
	\includegraphics[width=\columnwidth]{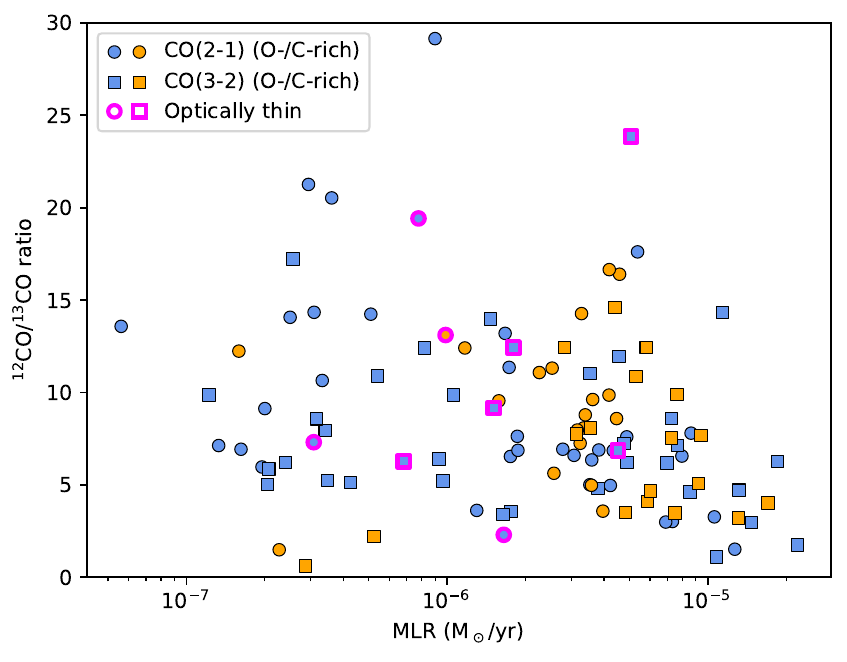}
    \caption{Plot of the $^{12}$CO/$^{13}$CO ratio as a function of mass-loss rate. Note that the optically thin C-rich IRAS 19008+0726 with a ratio of 69 and MLR of $4.6 \times 10^{-6}$~\msunyr\ has been excluded for clarity.}
    \label{fig:1213co_vs_MLR}
\end{figure}

Our data contains 80 sources with detections in both $^{12}$CO and $^{13}$CO so far.
We measure approximate $^{12}$CO/$^{13}$CO ratios by dividing the integrated intensities of both lines and multiplying by a factor of 0.874, equal to the frequency ratio cubed, to correct for differences in line strength (as is also done in, e.g., \citealp{de_beck_probing_2010}). In most cases this provides a lower limit on the ratio as most $^{12}$CO lines are optically thick and hence their integrated intensities do not linearly scale with abundance, unlike their optically thin counterparts. 
\citet{Saberi2020} examine this question in more detail, showing that chemical and excitation effects compound these optical-depth differences. This means that inference of reliable $^{12}$CO/$^{13}$CO isotopologue ratios requires the use of line radiative-transfer models. As a result, we limit our discussion here to observed line ratios, and do not attempt to directly infer the isotope ratios themselves, although in an ideal case these two parameters should be related.
The measured $^{12}$CO/$^{13}$CO ratios are plotted against MLR in Figure~\ref{fig:1213co_vs_MLR} (excluding the C-rich IRAS 19008+0726 which has a ratio of 69, MLR of $4.6 \times 10^{-6}$~\msunyr, and is optically thin).
We find a range of ratios from 0.62 to 69, with a mean value of 9.0 $\pm$ 0.7 and a median value of 7.3. For O-rich sources, we find slightly lower values, with a mean value of 8.6 $\pm$ 0.6 and a median of 6.9, while for C-rich sources we find slightly higher values, with a mean of 10 $\pm$ 2 and a median of 8.1. 
This is similar to the results from \citet{de_beck_probing_2010} who, for a sample of 27 sources, find a mean value of 10 $\pm$ 2 and a median value of 8.1 in both C-rich and O-rich sources. Limiting the analysis of our data to sources with optically thin $^{12}$CO lines should provide better estimates of the $^{12}$CO/$^{13}$CO ratios. We have visually identified ten sources (two of which are carbon-rich) where at least one of the $^{12}$CO lines shows a clear flat-topped or double-peaked profile, indicative of a low optical depth \citep{habing_asymptotic_2003}. 
An Anderson--Darling test \citep{anderson_asymptotic_1952} shows with high confidence ($p=0.001$) that the optically thin sources are systematically different (that is, their isotope ratios are drawn from a different underlying distribution) than the rest of the isotopic ratios.
The $^{12}$CO/$^{13}$CO ratios from these few optically thin lines have a range of 3 to 79, a mean value of 18 $\pm$ 6, and a median value of 10.5. 
This is slightly higher than the values from full sample, as expected, since the CO line ratio should be less underestimated than it is for optically thick lines, though we note that only having eleven sources limits the reliability of these statistics. 

We also compare our results to those of \citet{ramstedt_12co13co_2014} and \citet{milam_circumstellar_2009}, who use radiative-transfer modelling to determine abundances of $^{12}$CO and $^{13}$CO, and hence are better able to account for optical depth effects. \citet{ramstedt_12co13co_2014} have a sample of $\sim$60 stars, evenly split between oxygen-rich, carbon-rich, and S-type AGB stars. Overall they find a range of $^{12}$CO/$^{13}$CO ratios of 2 to 100, a mean value of 22 $\pm$ 3, and a median value of 17. The carbon-rich sources include the full range of ratio values, and have a larger mean value ($27 \pm 1$) but the same median value of 17. 
\citet{milam_circumstellar_2009} have a sample of 15 AGB stars, of which 11 are carbon-rich. The carbon-rich sources have a higher mean $^{12}$CO/$^{13}$CO ratio of $38 \pm 2$, with a median of 29; while the oxygen-rich sources have a mean value of $27 \pm 3$ but a higher median value of 32. 
Both studies find larger values than this paper, even when we limit ourselves to the optically thin sources, reflecting their mitigation of the optical depth effects on the CO line ratio.
Overall, across both our data and previous studies, carbon-rich sources are found to have somewhat higher $^{12}$CO/$^{13}$CO ratios, as expected from the repeated dredge-ups that are the cause of their carbon-rich nature \citep{karakas_2016_yields}, but the differences with oxygen-rich sources are not as significant as might be expected, indicating that other effects have a larger impact on these measurements. 

Comparing our $^{12}$CO/$^{13}$CO ratios with MLR, we find a weak negative correlation (Spearman correlation coefficient of $-0.4$) between isotopic ratio and MLR (Fig.~\ref{fig:1213co_vs_MLR}). This correlation is more likely due to higher MLR sources having more optically thick $^{12}$CO lines, which will decrease their $^{12}$CO/$^{13}$CO ratios. The large scatter -- i.e., the difference in isotopic ratios derived from the CO(2-1) and CO(3-2) lines for individual sources -- even at low MLRs, seems to indicate that optical depth has a stronger effect on the $^{12}$CO/$^{13}$CO ratio than the evolutionary factors such as dredge up.
In contrast, the full radiative-transfer analysis by \citet{ramstedt_12co13co_2014} found no evidence for such a correlation.

\subsection{Mass-loss rates}\label{sect:mlr}

Empirical gas MLRs were calculated using the formula from \citet{ramstedt_reliability_2008} (Equation~\ref{eq:ramstedt}), as described in more detail in Section~\ref{sect:data_analysis}.
There are significant uncertainties associated with this formula, mainly from the formula parameters themselves, which depend on the observed CO transition, but also from the measured input parameters such as integrated intensity and distance. The formula may also produce unreliable estimates at high MLRs \citep[$\dot{M} \geq 10^{-5}$~\msunyr;][]{de_beck_probing_2010}. Altogether, this results in uncertainties on the calculated MLRs of about an order of magnitude, but the aggregate results are still useful as a first approximation in the absence of full radiative-transfer modelling.

Calculated MLRs range from $1.9 \times 10^{-8} - 1.1 \times 10^{-4}$~\msunyr, whereas DPRs range from $1.2 \times 10^{-11} - 1.1 \times 10^{-6}$~\msunyr \citep{scicluna_nearby_2022}. 
We find that the MLRs tend to increase with the NESS tiers, as expected since the tiers' definition includes an increase in DPR, which is related to MLR. The median MLR values for the different tiers are: "low": $7.5 \times 10^{-8}$, "intermediate": $3.2 \times 10^{-7}$, "high": $3.2 \times 10^{-6}$, and "extreme": $8.4 \times 10^{-6}$ \msunyr. 
We also find differences between the O-rich and C-rich sources, as expected from AGB evolution.
For O-rich sources, the mean MLR is $2.7 \pm 0.2 \times 10^{-6}$ and the median MLR is $8.3 \times 10^{-7}$~\msunyr. The C-rich sources have slightly higher average MLRs, with a mean value of $7 \pm 1 \times 10^{-6}$ and a median of $4.0 \times 10^{-6}$~\msunyr. 

\begin{figure}
	\includegraphics[width=\columnwidth]{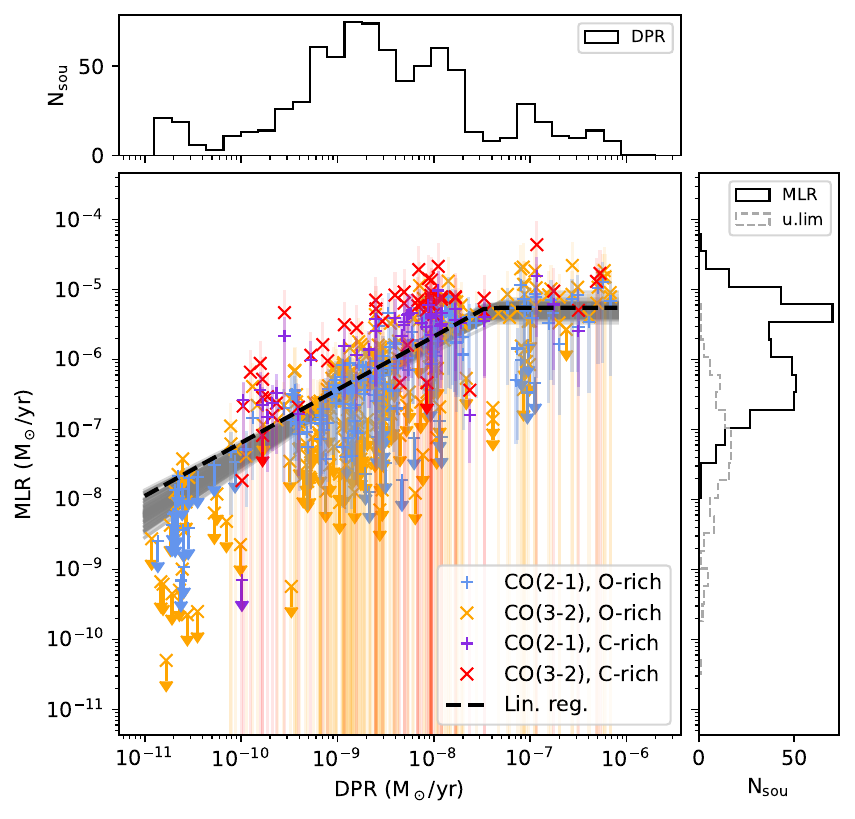}
    \caption{Plot of log(MLR) including error bars vs log(DPR) with a best-fit broken power law in dashed black. The grey lines show random draws from the posterior distribution, as an indication of the uncertainty in the fit. The marginal distributions for the MLR and DPR are shown as histograms. A histogram (dashed) is also shown for the cases where the MLRs are upper limits.}
    \label{fig:scatter_MLR_DPR}
\end{figure}

We find a best fit relating the MLR and DPR (including the upper limits) with a broken power-law distribution, using the Markov-Chain Monte Carlo (MCMC) implementation \texttt{EMCEE} \citep{foreman-mackey_emcee:_2013}, as shown in Figure~\ref{fig:scatter_MLR_DPR}. In log(MLR)-log(DPR) space, this straight line follows the equation
\begin{equation}\label{eq:mlr-dpr}
    \log{\left(\frac{\mathrm{MLR}}{\mathrm{M}_\odot\ \mathrm{yr}^{-1}}\right)} = 0.82^{+0.06}_{-0.06}\ \log{\left(\frac{\mathrm{DPR}}{\mathrm{M}_\odot\ \mathrm{yr}^{-1}}\right)} + 0.82^{+0.51}_{-0.52}
\end{equation}
until the break at log(DPR) = $-7.45^{+0.15}_{-0.13}$,
or DPR $\sim 3.5 \times 10^{-8}$~\msunyr. This "saturation value" for the MLR is at MLR = $5.2^{+1.0}_{-0.8} \times 10^{-6}$~\msunyr. 
The given uncertainties are the 68\% credible interval on each parameter, and a corner plot of the full MCMC distributions is shown in Figure~\ref{fig:app:MLRDPRcorner}.
Despite the large uncertainties on each data point and the inherent scatter, the fit is well constrained and clearly implies there is no single gas-to-dust ratio for AGB stars, as discussed further in Section~\ref{sect:gd}. We also note here that the DPR estimates are based on SED fits that only take the warm dust emission into account. In contrast, the MLR is based on CO emission, which integrates over larger radii in the circumstellar envelope. Therefore, the MLR and DPR may be probing material ejected at different epochs.

Interestingly, the position of the break in the power law coincides with a relative paucity of sources in both the MLR and DPR distributions, as seen in the histograms in Figure~\ref{fig:scatter_MLR_DPR}, at values which correspond to the high-DPR end of the "high" tier of NESS sources. We note this paucity of sources is also seen in the full NESS sample, so is not due to a bias in the data observed so far. It is possible this is an evolutionary effect, where the majority of low-mass stars have lower MLRs than $\sim$10$^{-6}$~\msunyr, whereas intermediate-mass stars will instead tend to cluster around the highest possible MLR, which is expected from the theory of dust-driven winds to be $\sim$10$^{-5} - 10^{-4}$~\msunyr \citep{lamers_introduction_1999}. It is also possible this is partly an observational effect, as the saturation of the CO lines for higher MLRs will make it more difficult to derive accurate MLRs from their integrated intensities.  Finally, it could also be a temporal or external effect, such as from coincidences in the timings of thermal pulses, or the impact of binarity. 

We find that MLRs calculated from CO (3--2) lines are systematically higher than MLRs from CO (2--1) lines, by a median factor of $\sim$1.8, though they also have larger uncertainties. 
A systematic offset of similar order was noted in \citet{de_beck_probing_2010}, with Eq.~\ref{eq:ramstedt} predicting higher MLR for higher-$J$ transitions (their figure 10). They explain this as due to differences in the handling of cooling between \citet{ramstedt_reliability_2008} and \citet{de_beck_probing_2010}. Better radiative transfer modelling of the CO lines is required to clarify the cause of this offset.

\subsection{Gas-to-dust ratios}\label{sect:gd}

\begin{figure}
	\includegraphics[width=\columnwidth]{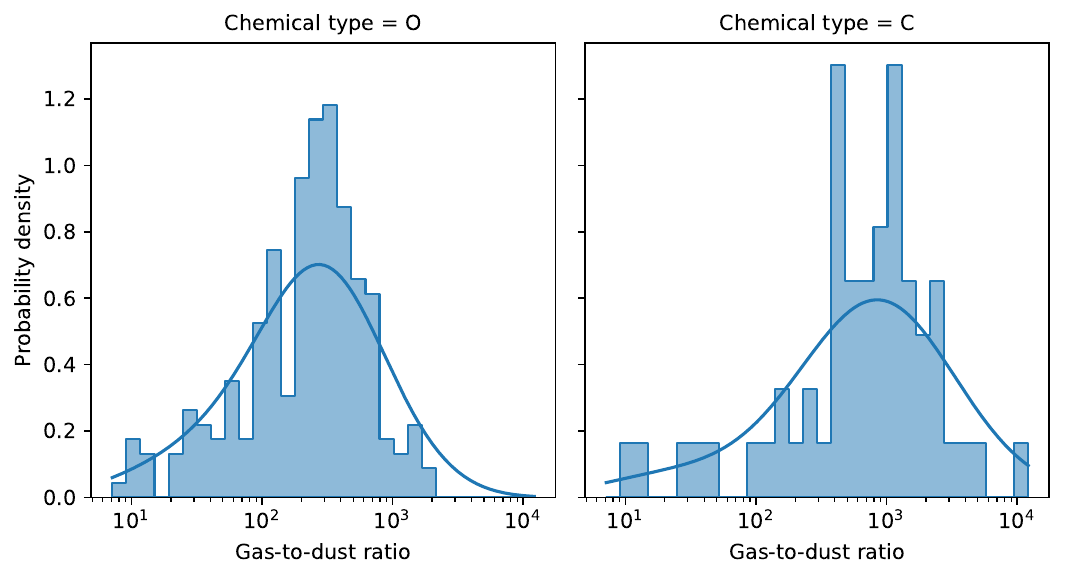}
    \caption{Histogram and KDE of the gas-to-dust ratios, for O-rich and C-rich sources, respectively.}
    \label{fig:hist_gasdust}
\end{figure}

Despite indications that there is not a constant AGB gas-to-dust ratio across the whole parameter space, it is still useful to characterise its distribution from our large sample using summary statistics. A histogram and KDE of the gas-to-dust ratios is shown in Figure~\ref{fig:hist_gasdust}.
Dividing the MLR by the DPR, we find a very large range of gas-to-dust ratios, between 2 and 16\,600. 
The extreme values of this range are probably not accurate, and we reiterate that there are large uncertainties on both the MLR and DPR values.
To quantify the average gas-to-dust ratio, we have taken the 16th, 50th, and 84th percentiles of the gas-to-dust ratio distribution to find a median value of 290, and a 68\% confidence interval ranging from 80 to 790. 
There are also large differences between the O-rich and C-rich sources. The O-rich sources have a median value of 250 and the 68\% confidence interval ranges from 70 to 580, while for C-rich sources the median value is 680 and the 68\% confidence interval ranges from 160 to 1990. 

Our gas-to-dust ratio estimates are higher than what has been found in the literature, but broadly consistent when the large uncertainties are taken into account. 
\citet{knapp_mass_1985-1} find gas-to-dust ratios of 160 for O-rich sources and 400 for C-rich sources in their sample of 40 AGB stars, as compared with 250 and 680 for our O-rich and C-rich sources, respectively. 
\citet{groenewegen_millimeter_1999} find a range of gas-to-dust ratios from 7.9 to 1570, with a median value of 152, for a sample of 72 sources, of which $\sim$60\% are O-rich; our overall median gas-to-dust value of 290 is about a factor of two higher, from a much larger sample that is $\sim$87\% O-rich.

We have also calculated non-parametric Spearman rank correlations between the gas-to-dust ratio and the MLR, DPR, and expansion velocity. We find a significant correlation only for the DPR: a negative correlation coefficient of $-0.53$ with the gas-to-dust ratio. Given that the MLR and DPR are correlated, the lack of correlation between MLR and gas-to-dust ratio even though gas-to-dust ratio \emph{is} correlated with DPR is unexpected. Why should these two correlations cancel out so effectively? One possibility is that as the density at the base of the wind increases, more dust condenses. This then translates into a higher radiative momentum available to accelerate the wind; however, the sub-linear scaling of MLR with DPR shows that this increased momentum does not translate directly into more efficient mass loss. This in turn explains the increase in velocity at higher MLR (see below); as the condensation efficiency increases, a larger fraction of momentum must be translated to velocity as the gas-to-dust ratio decreases. In contrast with our results, neither \citet{knapp_mass_1985-1} nor \citet{groenewegen_millimeter_1999} find any significant correlation between gas-to-dust ratio and MLR or DPR. 

The power-law relationship between MLR and DPR found in Section~\ref{sect:mlr} implies that the gas-to-dust ratio decreases with increasing MLR and DPR, such that sources with higher mass-loss rates have a wind with a larger proportion of dust. This is broadly consistent with the scenario of a dust-driven wind. While the C-rich sources follow this trend for higher gas-to-dust ratio at higher MLR in Figure \ref{fig:scatter_MLR_DPR}, the population is offset towards higher mass-loss rates. This could be a result of the fact that carbon-rich dust is more efficient at driving the wind, and therefore less dust is required to achieve a higher MLR. However, it is possible that the systematics in assumptions in the modelling of the dust SED contribute to this difference.

\subsection{Comparison with combined literature sample}\label{sect:comparison_lit}

\begin{figure}
	\includegraphics[width=\columnwidth]{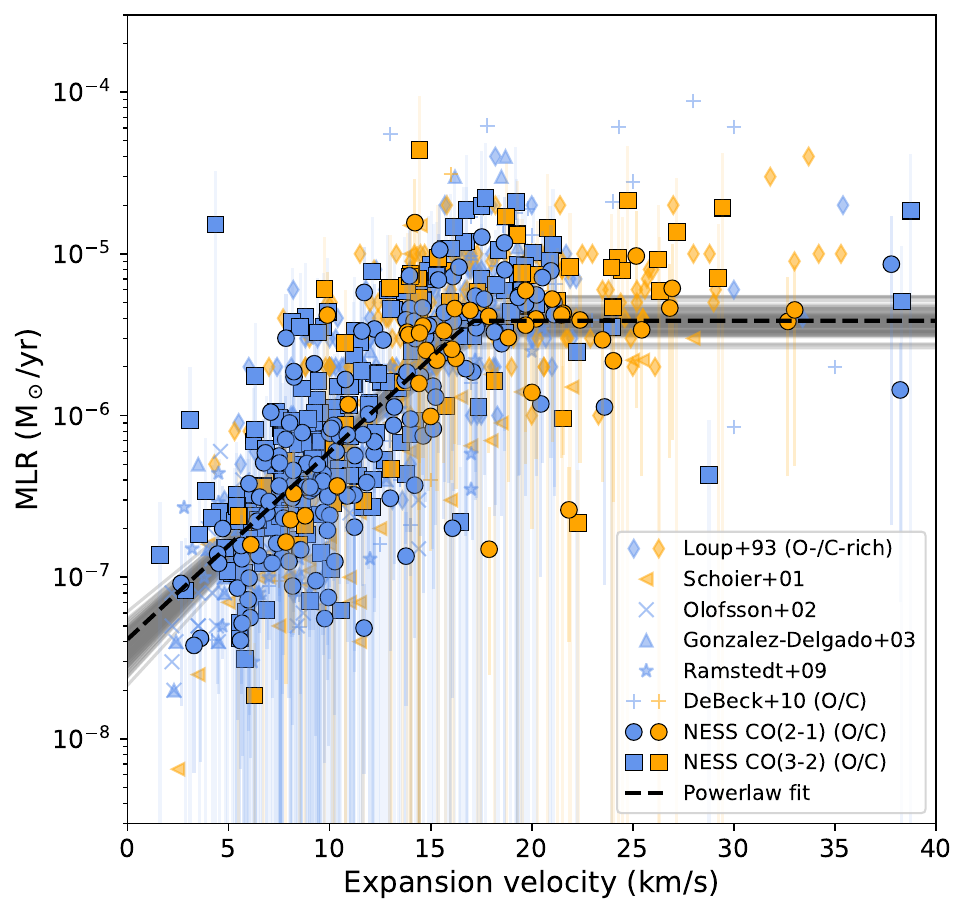}
    \caption{MLR (including error bars) vs expansion velocity, compared with samples from literature. Oxygen-rich (and S-type) sources are plotted in blue while carbon-rich sources are in orange. A broken power law fit to the NESS data is shown in dashed black, and the grey lines show random draws from the posterior distribution as an indication of the uncertainty in the fit.}
    \label{fig:MLRvsvinf_lit}
\end{figure}

We compare our empirical MLRs and expansion velocities against a literature sample, which combines the results of \citet{loup_co_1993,schoier_models_2001,olofsson_mass_2002,gonzalez_delgado_thermal_2003,ramstedt_circumstellar_2009}; and \citet{de_beck_probing_2010}. These studies are described in Appendix~\ref{sect:lit_sample} and plotted against the NESS results in Figure~\ref{fig:MLRvsvinf_lit}. Overall, our results are similar to the literature results, mainly probing sources with MLR $\sim 10^{-7} - 10^{-5}$~\msunyr\ and expansion velocities $\sim$5--20~\kms, with some outliers at higher velocities. The current NESS data shows some dearth of low-MLR and low-velocity sources, and fewer of the high-velocity outliers, but also shows some excess of lower-MLR sources at all velocities and some high-MLR outliers. Some of these outliers are most likely not AGB stars, for example the O-rich source with MLR $\sim 10^{-5}$~\msunyr\ and velocity $\sim$ 5~\kms\ is IRAS~19597+3327A: an infrared source that hasn’t previously been studied in CO with a very bright but narrow CO line. However, others seem to be AGB stars that have never been observed in CO before; for instance, the C-rich IRAS~00084-1851 (AC Cet), which has a very low MLR of $\sim 2 \times 10^{-8}$~\msunyr\ and an expansion velocity around 6~\kms, is classified as a long-period variable \citep{samus_general_2017}. There are only two additional C-rich stars in the NESS data that have not been previously observed in CO: IRAS 17565-2035 and IRAS 19321+2757 (IRC +30374).

We have used the same MCMC implementation as in Section~\ref{sect:mlr} to fit a broken power law to the MLR as a function of expansion velocity, as shown in Figure~\ref{fig:MLRvsvinf_lit}. We find
\begin{equation}\label{eq:mlr-vinf}
    \log{\left(\frac{\mathrm{MLR}}{\mathrm{M}_\odot\ \mathrm{yr}^{-1}}\right)} = 0.120^{+0.008}_{-0.008} \left(\frac{v_\mathrm{inf}}{\mathrm{km\ s}^{-1}}\right) - 7.4^{+0.9}_{-0.9}
\end{equation}
until the break at $v_\mathrm{inf} = 16.8^{+0.8}_{-0.7}$~\kms. 
This corresponds to an MLR saturation value of $3.9^{+0.5}_{-0.4} \times 10^{-6}$~\msunyr. The given uncertainties are the 68\% credible interval on each parameter, and a corner plot of the full MCMC distributions is shown in Figure~\ref{fig:app:MLRvinfcorner}.
To determine whether this slope is due entirely to the dependence of MLR on expansion velocity in Eq.~\ref{eq:ramstedt}, we perform a $t$-test with the null hypothesis equal to the slope expected given the values of $b_J$ in \citet{ramstedt_reliability_2008}. The estimates for the slopes obtained are around 2.3. The $p$-value of the null hypothesis is $\sim 10^{-23}$, clearly rejected at high significance.

There are only 18 NESS sources with $v_{\rm inf} > 23$ km s$^{-1}$, of which 12 are carbon stars with velocities up to $33$~\kms. Except for one O--rich Mira ($v_{\rm inf}\approx 24$~\kms), the remaining are either RSGs or post-AGB/binary stars, or they show clear deviations from the soft-parabola profile in at least one of the two lines. OH/IR stars in our sample are restricted to $v_{\rm inf}<22$~\kms. These results are consistent with the expectation from hydrodynamic models -- single M-type models tend to have $v_{\rm inf} < 25-30$ km s$^{-1}$, while single C-star models extend beyond 30 km s$^{-1}$ \citep{bladh_carbon_2019, bladh_extensive_2019}. Our results are consistent with those of \citet{de_beck_probing_2010} -- once we eliminate RSGs, sources that have evolved beyond the AGB, and sources with line profiles that deviate from a soft parabola, the highest expansion velocity in their sample (their Table A.1) belongs to an OH/IR star (25~\kms).\\
 
\begin{figure}
	\includegraphics[width=\columnwidth]{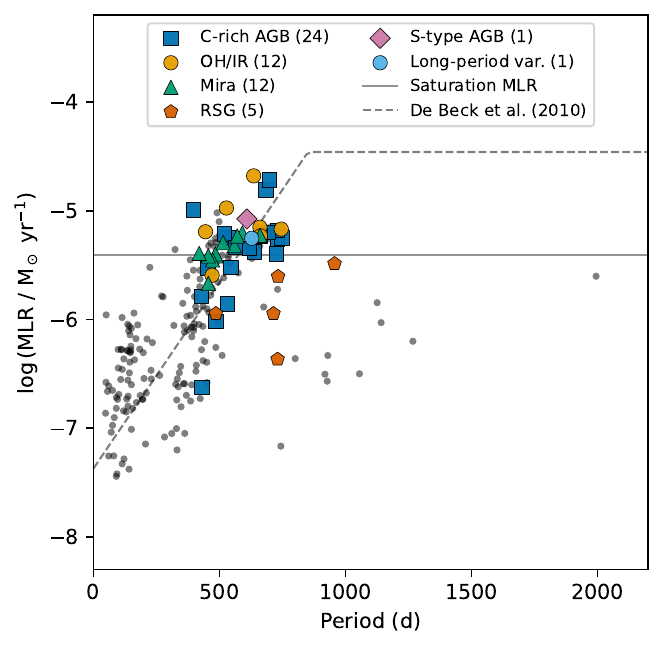}
    \caption{Mean MLR vs median period for our sample (grey circles). Sources with $v_{\rm inf} > 17$~\kms are coloured according to their SIMBAD object type. The fit from \citet{de_beck_probing_2010} (dashed curve) reproduces the increasing trend in our data well. The saturation MLR from our broken-power law fit (Equation~\ref{eq:mlr-vinf}, solid line) is also shown for comparison.}
    \label{fig:MLR_vs_period}
\end{figure}

\citet{vassiliadis_evolution_1993} noted that the MLR for Galactic Miras increases exponentially with pulsation period for periods $\lesssim 500$ d. Beyond this period, the MLR seems to saturate at values consistent with the superwind phase, where the MLR is at the radiation pressure limit given by $L/(cv_\mathrm{inf})$ with $v_\mathrm{inf}\approx 15$ \kms, which is comparable to the location of the knee ($v_{\rm inf}\approx 17$~\kms) in Figure~\ref{fig:MLRvsvinf_lit}. \citet{de_beck_probing_2010} found a similar trend, with a somewhat higher period before saturation ($\sim 850$ d, their Figure 14). In Figure~\ref{fig:MLR_vs_period}, we plot our mean MLRs (averaged over both CO lines) against the median values calculated from the pulsational periods compiled by \citet{McDonald2025_NessCat} for the NESS sample. Sources with $v_{\rm inf} >17$ km s$^{-1}$, corresponding to the break found in the MLR vs. $v_{\rm inf}$ plot, are highlighted in the figure and grouped by SIMBAD object type. Our data replicate the increasing trend followed by saturation around $\sim 750$ d. The rising trend of MLR with expansion velocity at low MLRs is well reproduced by the relation from \citet{de_beck_probing_2010} (solid line in the figure). The MLR saturation values quoted by \citet{vassiliadis_evolution_1993} and \citet{de_beck_probing_2010} are roughly an order of magnitude higher than our value. This difference is not unexpected, given the scatter in their data, the tendency of the \citet{ramstedt_reliability_2008} relation to underestimate high MLRs, and the bias of literature samples toward higher-MLR stars (see below).\\
To compare our results with the literature sample, we have divided the NESS results into two subsamples: the 221 sources that are part of the literature sample (NESS-lit) and the 272 that are not (NESS-nonlit). Both subsamples have similar median MLR values: $1.7 \times 10^{-6}$ M$_\odot$\,yr$^{-1}$ and $7.5 \times 10^{-7}$ M$_\odot$\,yr$^{-1}$ for NESS-lit and NESS-nonlit, respectively.
The range of MLRs is also comparable between NESS-lit ($3.1 \times 10^{-8} - 4.4 \times 10^{-5}$ M$_\odot$\,yr$^{-1}$) and NESS-nonlit ($1.9 \times 10^{-8} - 1.9 \times 10^{-5}$ M$_\odot$\,yr$^{-1}$). However, the NESS-lit subsample extends to larger expansion velocities (38.8 \kms) than NESS-nonlit (26.9 \kms), and has a higher median expansion velocity of 13.4 \kms, as compared to a median expansion velocity of 10.3 \kms\ in the NESS-nonlit subsample.

\begin{figure}
	\includegraphics[width=\columnwidth]{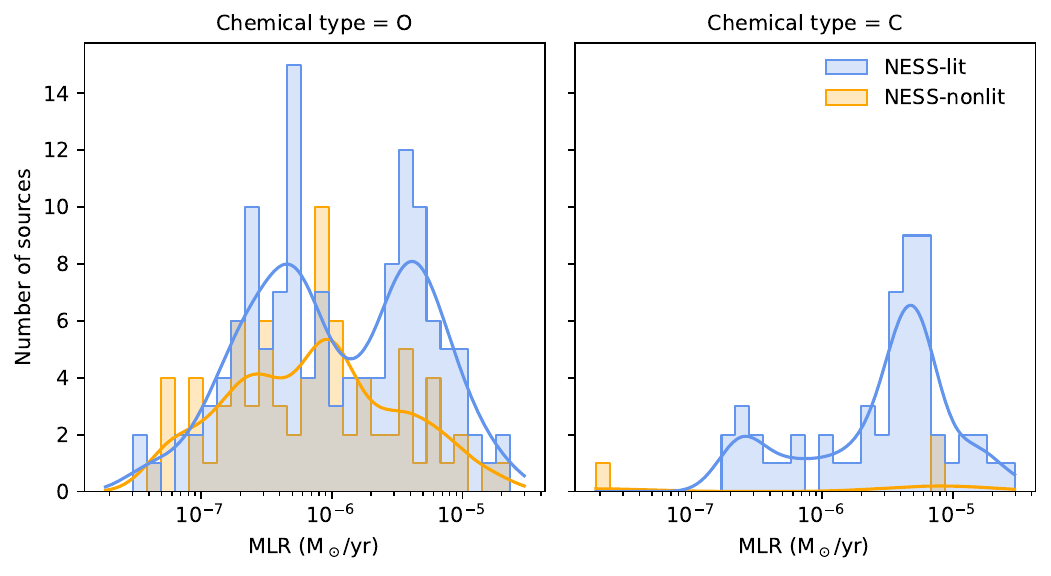}
    \caption{Histogram and KDE of the MLR derived for NESS sources which are in the literature sample (NESS-lit), and those which are not (NESS-nonlit).}
    \label{fig:lit_hist_MLR}
\end{figure}

In the MLR histogram in Figure~\ref{fig:lit_hist_MLR},
we can see that overall higher MLR sources are over-represented in the NESS-lit subsample. This is expected, as the literature sample is drawn from previously detected sources and hence biased towards brighter and higher-MLR stars.
There are also stark differences in chemical composition and detection rates between the two subsamples, as seen in Table~\ref{tab:detec_vs_lit}. The NESS-nonlit subsample contains only three C-rich sources, while the other 95\% of the C-rich sources in our data belong to the NESS-lit subsample. Most (60\%) of the sources in the NESS-nonlit sample with uncertain chemical classifications are in the low tier, whereas the known carbon stars from the NESS-lit sample are typically found in the high and intermediate tiers. The uncertain classifications therefore do not significantly alter our conclusions. Furthermore, for NESS-lit, 97\% of sources are detected in CO (2--1) and 90\% are detected in CO (3--2), as compared with only $\sim$60\% in both lines for the NESS-nonlit subsample. Many of the detected NESS-nonlit sources have no previously published CO data. Another difference is the proportion of sources with a soft-parabola-shaped CO line: 80\% of the NESS-lit sources but nearly 100\% of the NESS-nonlit sources, showing that unusual sources are over-represented in the literature sample.  
This shows the value in building a volume-limited sample: NESS includes a lot of under-studied sources, forming a more complete picture of the local population of AGB stars and, while many of these observations are non-detections and hence provide only upper limits, this is still a vital aspect of characterizing the entire AGB population, especially given our homogeneous observing setup. 

% \begin{table*}[ht]
% \caption{Detection statistics for the NESS data in total, and divided into two subsamples: the sources in the literature sample (NESS lit), and the sources not in the literature sample (NESS non-lit).}
% \label{tab:detec_vs_lit} 
% \centering 
% \begin{tabular}{rrcccccc}
% \hline\hline
% & Sample: & All NESS & \% & NESS lit & \% & NESS non-lit & \% \\
% \hline
% Number of sources & Total & 485 & 100\% & 217 & 45\% & 268 & 55\% \\
%  & O-rich & 421 & 100\% & 156 & 37\% & 265 & 63\% \\
%  & C-rich & 64 & 100\% & 61 & 95\% & 3 & 5\% \\
% CO(2-1) detections & Total & 210/259 & 81\% & 147/151 & 97\% & 63/108 & 58\% \\
%  & O-rich & 168/216 & 78\% & 106/110 & 96\% & 62/106 & 58\% \\
%  & C-rich & 42/43 & 98\% & 41/41 & 100\% & 1/2 & 50\% \\
% $^{13}$CO(2-1) detections & Total & 57/136 & 42\% & 57/123 & 46\% & 0/13 & 0\% \\
%  & O-rich & 36/94 & 38\% & 36/82 & 44\% & 0/12 & 0\% \\
%  & C-rich & 21/42 & 50\% & 21/41 & 51\% & 0/1 & 0\% \\
% CO(3-2) detections & Total & 320/428 & 75\% & 164/183 & 90\% & 156/245 & 64\% \\
%  & O-rich & 268/374 & 72\% & 115/132 & 87\% & 153/242 & 63\% \\
%  & C-rich & 52/54 & 96\% & 49/51 & 96\% & 3/3 & 100\% \\
% $^{13}$CO(3-2) detections & Total & 55/178 & 31\% & 54/126 & 43\% & 1/52 & 2\% \\
%  & O-rich & 37/147 & 25\% & 36/95 & 38\% & 1/52 & 2\% \\
%  & C-rich & 18/31 & 58\% & 18/31 & 58\% & 0/0 & -- \\
% Soft parabola shaped & Total & 432/485 & 89\% & 169/217 & 78\% & 263/268 & 98\% \\
%  & O-rich & 387/421 & 92\% & 125/156 & 80\% & 262/265 & 99\% \\
%  & C-rich & 45/64 & 70\% & 44/61 & 72\% & 1/3 & 33\% \\
% \hline
% \end{tabular} 
% \end{table*}

\begin{figure}
	\includegraphics[width=\columnwidth]{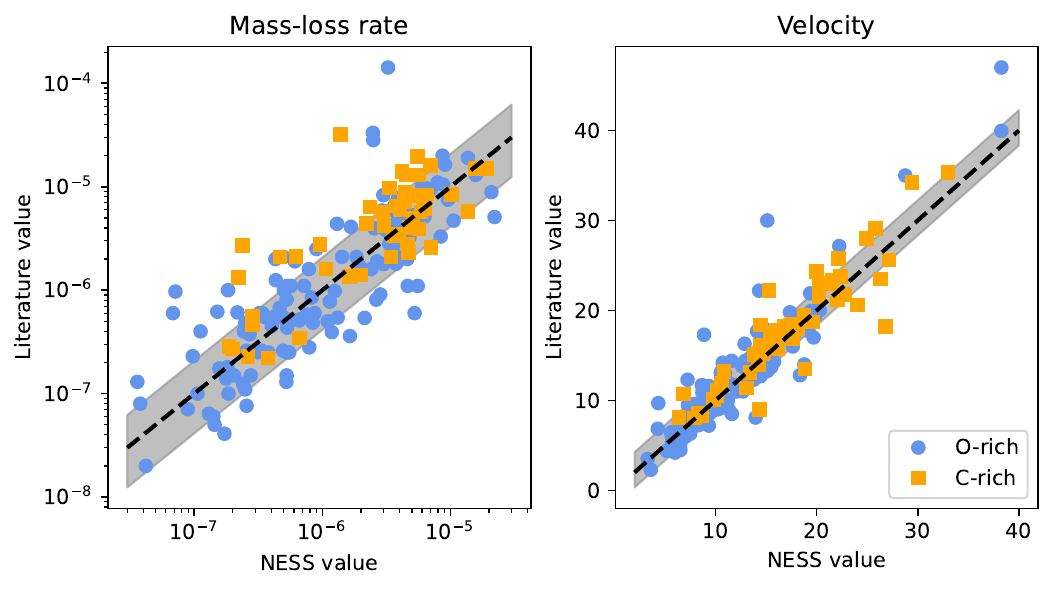}
    \caption{Mass-loss rates and expansion velocities derived in this paper, as compared with the values from literature, for sources common to both samples. The dashed line corresponds to a ratio of 1 and the shaded region covers  the central 68\% of values.}
    \label{fig:hist_NESSvslit}
\end{figure}

We also compare our derived MLR and expansion velocity values with those calculated in the literature papers described in Appendix~\ref{sect:lit_sample} for the 162 individual sources with CO detections in both NESS and the literature sample. Dividing our MLR with the mean of literature values for each source yields a range between $0.02 - 8.7$ with a median ratio of 0.88, while for the expansion velocity we find a range between $0.5 - 1.7$ with a median ratio of 0.96, shown in Figure~\ref{fig:hist_NESSvslit}. 
While this spread is large, the empirical MLR calculations have uncertainties of about an order of magnitude, and most values are within these uncertainties. We also note that the literature MLRs are calculated in a wide variety of ways (see Appendix~\ref{sect:lit_sample}), for which uncertainties are not always quantified. 
The ratio of the expansion velocities has a narrower distribution, centred on unity as expected, though some ratios show discrepancies of up to a factor $\sim$2. Some of the literature observations are of CO (1--0) or SiO lines, rather than CO (2--1) or (3--2), which may explain some of the discrepancy; for example, the higher excitation of SiO lines may result in a lower velocity since it tends to be emitted in the inner wind, while CO(1--0) may produce broader lines since it probes the outermost parts of the envelope. Furthermore, low signal-to-noise observations can underestimate line widths. We note that, of the spectra in our dataset with inferred expansion velocities outside the 68$^\text{th}$ percentile, almost 80\% have soft parabola shapes. This indicates that the NESS velocity estimates for these spectra are reliable despite their disagreement with the literature values.

\section{Conclusions}\label{sect:conclusions}
This paper has presented initial CO results from the NESS survey, observed and analysed in a homogeneous way using a new JCMT data reduction pipeline. We have demonstrated the advantages of a volume-limited sample, like NESS, for probing a large range of CO mass-loss rates. This first data release contains CO observations for 485 sources which are divided into four tiers with increasing distance and dust production rate (DPR).

We summarize our findings as follows:
\begin{itemize}
    \item We find overall detection rates of 81\% for CO~(2--1) and 75\% for CO~(3--2), including 59 sources with no previously published CO detections.
    \item 82\% of CO lines conform to a soft parabola shape, while 11 sources show a double wind. The majority of these double-wind sources are oxygen-rich, and tend to have lower-than-average mass-loss rates around a few $\times 10^{-7}$~\msunyr.
    \item Estimated $^{12}$CO/$^{13}$CO ratios have a median of 7.3 for the full sample and a median of 10.5 for the few sources where the $^{12}$CO line appears to be optically thin. Carbon-rich sources have overall slightly higher values than oxygen-rich, but the small differences indicate that other effects such as optical depth has a larger impact on the estimated ratios. We also find a weak negative correlation between the $^{12}$CO/$^{13}$CO ratio and mass-loss rate, which is also likely due to optical-depth effects.
    \item We calculate gas mass-loss rates (MLRs) using the empirical formula from \citet{ramstedt_reliability_2008}, resulting in uncertainties of about an order of magnitude. Overall, these estimates are similar to values found in literature and from models.
    \item We find a power-law relation between the MLR and DPR, up to a MLR saturation value of $5.3^{+1.0}_{-0.8} \times 10^{-6}$~\msunyr, implying there is no single gas-to-dust ratio for the population of AGB stars. 
    \item We show the distributions of gas-to-dust ratios for both oxygen-rich and carbon-rich AGB stars, which have median values of 250 and 680, respectively. The gas-to-dust ratio is found to be negatively correlated with the DPR, indicating that the dust-production process is more efficient at higher DPR, lowering the gas:dust ratio. While this correlation is at least in part due to the definition of the gas:dust ratio in terms of the MLR and DPR, the lack of correlation with MLR may indicate a change in the distribution of radiative momentum towards greater acceleration at higher DPR, explaining the increase in velocity at higher DPR. 
    \item We find a power-law relationship between MLR and expansion velocity, up to a MLR saturation value of $3.9^{+0.5}_{-0.4} \times 10^{-6}$~\msunyr, which corresponds to a velocity of $\sim$17~\kms. This is similar to the MLR saturation value found for the MLR-DPR relation, though the two values are not within each others credible intervals. 
    \item Comparing the NESS results with a large combined literature sample finds high mass-loss-rate sources are over-represented in the literature sample, especially among carbon-rich sources. The literature sources also have higher expansion velocities. 
    \item The most striking difference between the NESS results and the literature sample are the detection rates. Over 90\% of sources that are in the literature sample are detected in our data, while only $\sim$60\% of sources not in the literature sample are detected, and many of these have no previously published CO data. The proportion of sources showing a soft parabola line shape also differ: about 80\% of sources in the literature sample show a soft parabola shape compared with over 98\% of sources not in the literature sample. These statistics reflect the under-representation of low-MLR sources  and over-representation of extreme or unusual sources in literature samples. NESS detects significant numbers of low-MLR sources despite only being designed to sample them in a small volume, reflecting a potentially large bias in the literature.
    \item We also compare the calculated MLRs for individual sources with literature values, which are found be consistent within the (large) uncertainties. 
\end{itemize}

Overall the initial analysis of 485 NESS sources highlights the importance of our volume-limited approach in characterizing the local AGB population as a whole, and of including upper limits derived from non-detections, especially given our homogeneous observing strategy. As illustrated by the discrepancy in detection rates between the NESS sources included in and excluded from the literature sample, NESS is probing more low-MLR and under-observed sources. 

\section*{Acknowledgements}
SHJW acknowledges support from the Research Foundation Flanders (FWO) through grant 1285221N, from the ERC consolidator grant 646758 AEROSOL, from the Ministry of Science and Technology of Taiwan under grants MOST104-2628-M-001-004-MY3 and MOST107-2119-M-001-031-MY3, and from Academia Sinica under AS-IA-106-M03.
SS acknowledges support from UNAM-PAPIIT Programs IA104822 and IA104824.
FK acknowledges support from the Spanish Ministry of Science, Innovation and Universities, under grant number PID2023-149918NB-I00. This work was also partly supported by the Spanish program Unidad de Excelencia María de Maeztu CEX2020-001058-M, financed by MCIN/AEI/10.13039/501100011033.
TD is supported in part by the Australian Research Council through a Discovery Early Career Researcher Award (DE230100183).
 M.M. and R.W. acknowledge support from the STFC Consolidated grant (ST/W000830/1).
JH thanks the support of NSFC project 11873086. 
HK thanks the support of the National Research Foundation of Korea (NRF) grant (RS-2021-NR058398) and the Korea Astronomy and Space Science Institute (KASI) grant (Project No. 2025184102), both funded by the Korean Government (MSIT).
JPM acknowledges research support by the National Science and Technology Council of Taiwan under grant NSTC 112-2112-M-001-032-MY3.

The James Clerk Maxwell Telescope is operated by the East Asian Observatory on behalf of The National Astronomical Observatory of Japan; Academia Sinica Institute of Astronomy and Astrophysics; the Korea Astronomy and Space Science Institute; Center for Astronomical Mega-Science (as well as the National Key R\&D Program of China with No. 2017YFA0402700). Additional funding support is provided by the Science and Technology Facilities Council of the United Kingdom and participating universities in the United Kingdom and Canada. This paper made use of JCMT observations under program IDs M17BL002 and M20AL014.
The James Clerk Maxwell Telescope has historically been operated by the Joint Astronomy Centre on behalf of the Science and Technology Facilities Council of the United Kingdom, the National Research Council of Canada and the Netherlands Organisation for Scientific Research.
The Starlink software (Currie et al. 2014) is currently supported by the East Asian Observatory.
This research used the Canadian Advanced Network For Astronomy Research (CANFAR) operated in partnership by the Canadian Astronomy Data Centre and The Digital Research Alliance of Canada with support from the National Research Council of Canada the Canadian Space Agency, CANARIE and the Canadian Foundation for Innovation.
This work is sponsored (in part) by the Chinese Academy of Sciences (CAS), through a grant to the CAS South America Center for Astronomy (CASSACA) in Santiago, Chile.

This work has made use of Python packages 
Pyspeckit \citep{ginsburg_pyspeckit_2022},
PyVO \citep{GrahamPyVO2014},
Astropy \citep{robitaille_astropy_2013, price-whelan_astropy_2018, price-whelan_astropy_2022}, 
SciPy \citep{virtanen_scipy_2020}, 
pandas \citep{mckinney_data_2010}, 
NumPy \citep{harris_array_2020}, 
and Matplotlib \citep{hunter_matplotlib_2007}. This research has made use of the SIMBAD \citep{Wenger2000} database,
operated at CDS, Strasbourg, France.
This research has made use of NASA's Astrophysics Data System Bibliographic Services

\section*{Data Availability}
The data underlying this article are available in the article and in its online supplementary material, and on the NESS website \url{https://evolvedstars.space}.

%%%%%%%%%%%%%%%%%%%%%%%%%%%%%%%%%%%%%%%%%%%%%%%%%%



\bibliographystyle{aa}
\bibliography{papers, papers2} % if your bibtex file is called example.bib

\noindent\hrulefill

%%%%%%%%%%%%%%%%%%%%%%%%%%%%%%%%%%%%%%%%%%%%%%%%%%

%%%%%%%%%%%%%%%%% APPENDICES %%%%%%%%%%%%%%%%%%%%%

\clearpage
\onecolumn

\appendix

\section{Additional plots}
\subsection{Detection statistics}
\begin{table*}[ht]
\caption{Detection statistics for the NESS data in total, and divided into two subsamples: the sources in the literature sample (NESS lit), and the sources not in the literature sample (NESS non-lit).}
\label{tab:detec_vs_lit} 
\centering 
\begin{tabular}{rrcccccc}
\hline\hline
& Sample: & All NESS & \% & NESS lit & \% & NESS non-lit & \% \\
\hline
Number of sources & Total & 485 & 100\% & 217 & 45\% & 268 & 55\% \\
 & O-rich & 421 & 100\% & 156 & 37\% & 265 & 63\% \\
 & C-rich & 64 & 100\% & 61 & 95\% & 3 & 5\% \\
CO(2-1) detections & Total & 210/259 & 81\% & 147/151 & 97\% & 63/108 & 58\% \\
 & O-rich & 168/216 & 78\% & 106/110 & 96\% & 62/106 & 58\% \\
 & C-rich & 42/43 & 98\% & 41/41 & 100\% & 1/2 & 50\% \\
$^{13}$CO(2-1) detections & Total & 57/136 & 42\% & 57/123 & 46\% & 0/13 & 0\% \\
 & O-rich & 36/94 & 38\% & 36/82 & 44\% & 0/12 & 0\% \\
 & C-rich & 21/42 & 50\% & 21/41 & 51\% & 0/1 & 0\% \\
CO(3-2) detections & Total & 320/428 & 75\% & 164/183 & 90\% & 156/245 & 64\% \\
 & O-rich & 268/374 & 72\% & 115/132 & 87\% & 153/242 & 63\% \\
 & C-rich & 52/54 & 96\% & 49/51 & 96\% & 3/3 & 100\% \\
$^{13}$CO(3-2) detections & Total & 55/178 & 31\% & 54/126 & 43\% & 1/52 & 2\% \\
 & O-rich & 37/147 & 25\% & 36/95 & 38\% & 1/52 & 2\% \\
 & C-rich & 18/31 & 58\% & 18/31 & 58\% & 0/0 & -- \\
Soft parabola shaped & Total & 432/485 & 89\% & 169/217 & 78\% & 263/268 & 98\% \\
 & O-rich & 387/421 & 92\% & 125/156 & 80\% & 262/265 & 99\% \\
 & C-rich & 45/64 & 70\% & 44/61 & 72\% & 1/3 & 33\% \\
\hline
\end{tabular} 
\end{table*}
\begin{figure}[!ht]
    \centering
    \includegraphics[width=0.75\columnwidth]{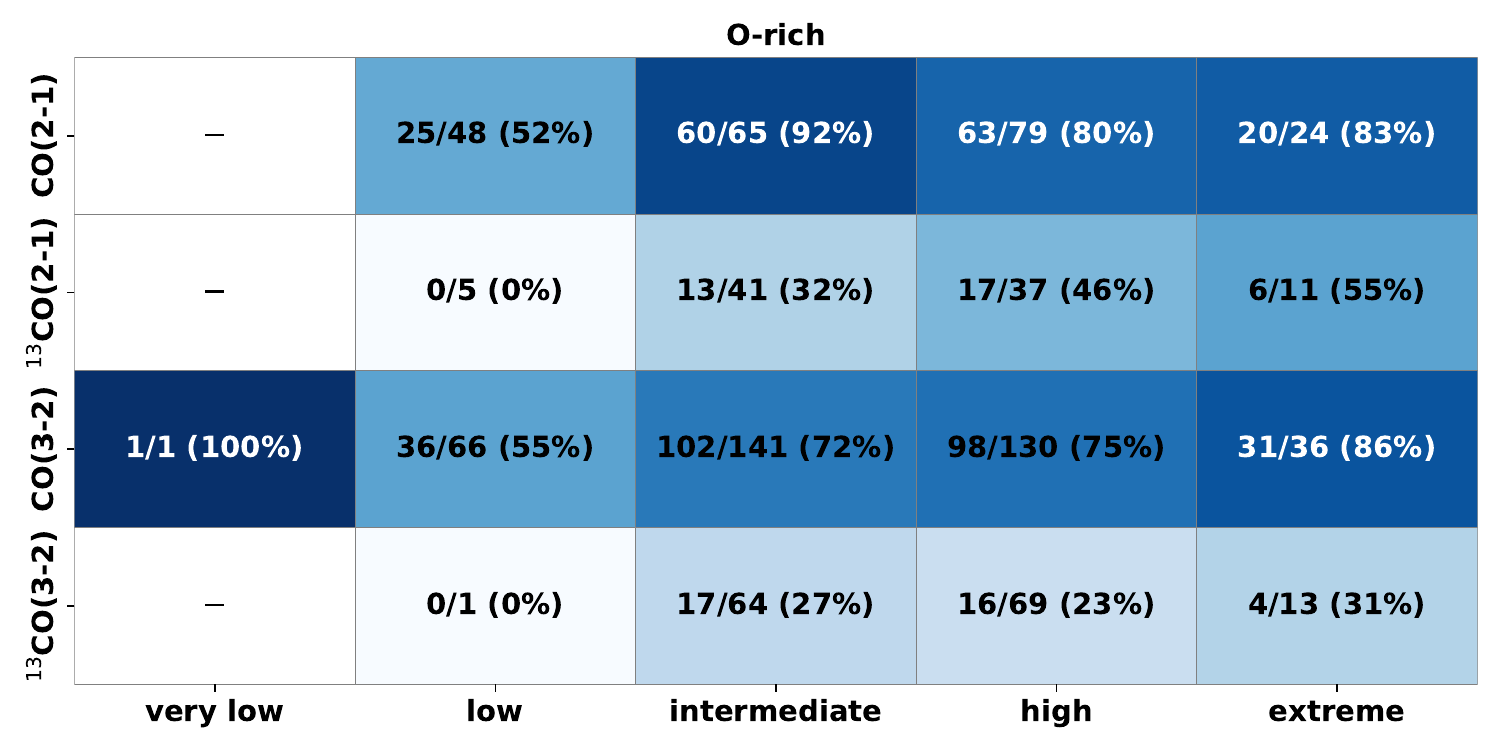}\\
    \includegraphics[width=0.75\columnwidth]{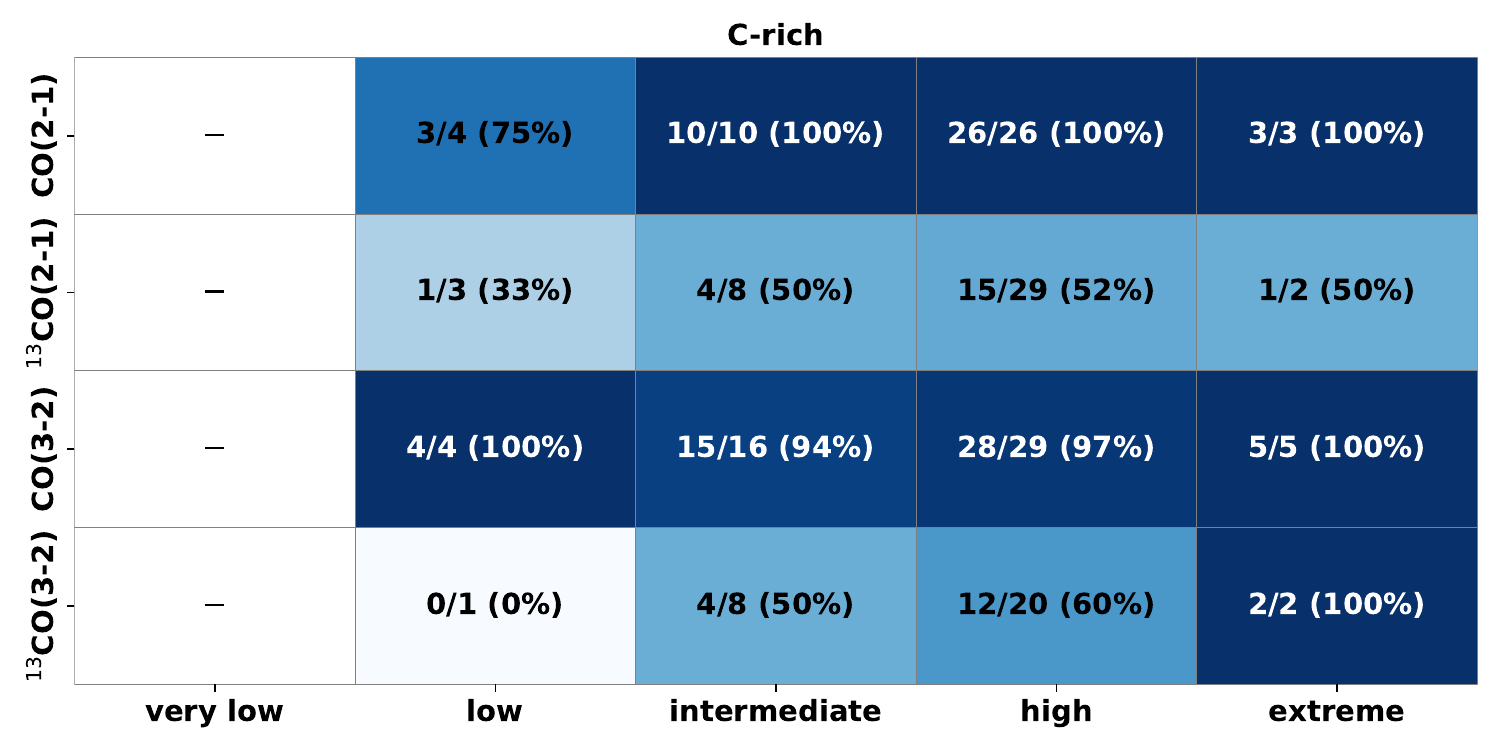}\\
    \includegraphics[width=0.75\columnwidth]{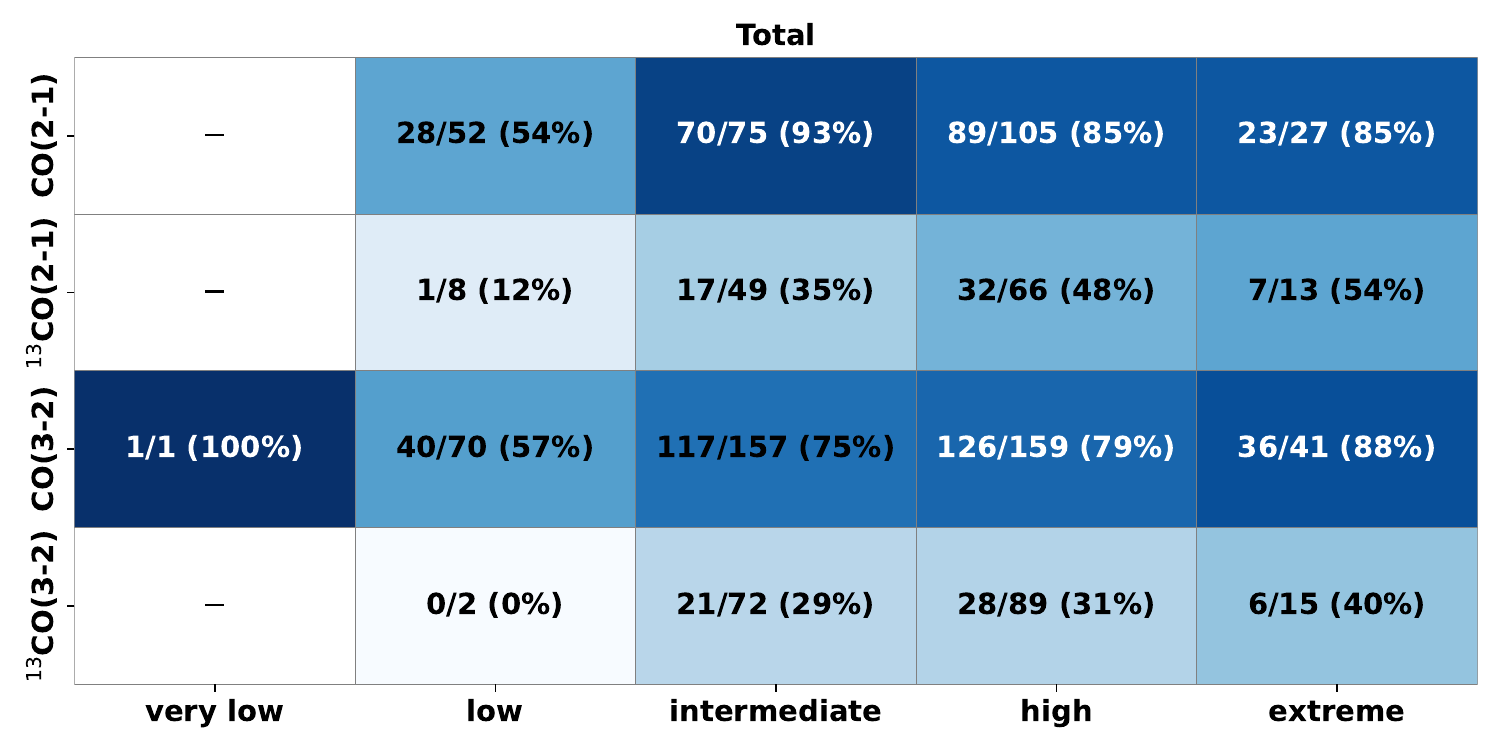}
    \caption{Heatmaps showing detections in the JCMT heterodyne data processed in this paper for the O--rich (top), C--rich (centre), and the full sample of observed NESS sources.}
    \label{fig:detection-stats}
\end{figure}

\clearpage

\subsection{Corner plots}
\begin{figure}[ht]
	\includegraphics[width=0.5\columnwidth]{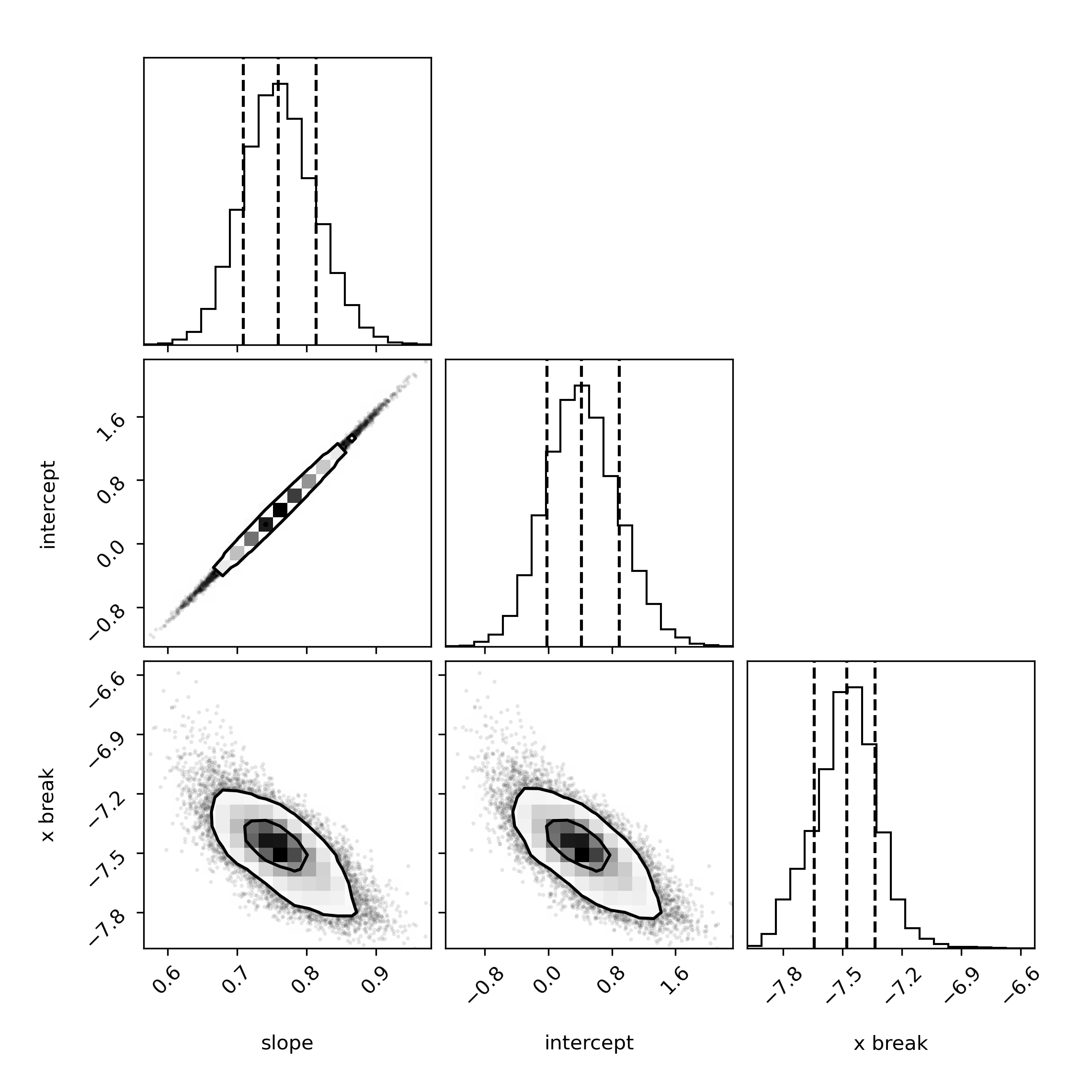}
    \caption{Corner plot of the MCMC fit to the log(MLR)-log(DPR) plot in Figure~\ref{fig:scatter_MLR_DPR}. The 16th, 50th, and 84th percentiles are marked with dashed lines, and the contour plots show 1$\sigma$ and 2$\sigma$ contours.}
    \label{fig:app:MLRDPRcorner}
\end{figure}

\begin{figure}[ht]
	\includegraphics[width=0.5\columnwidth]{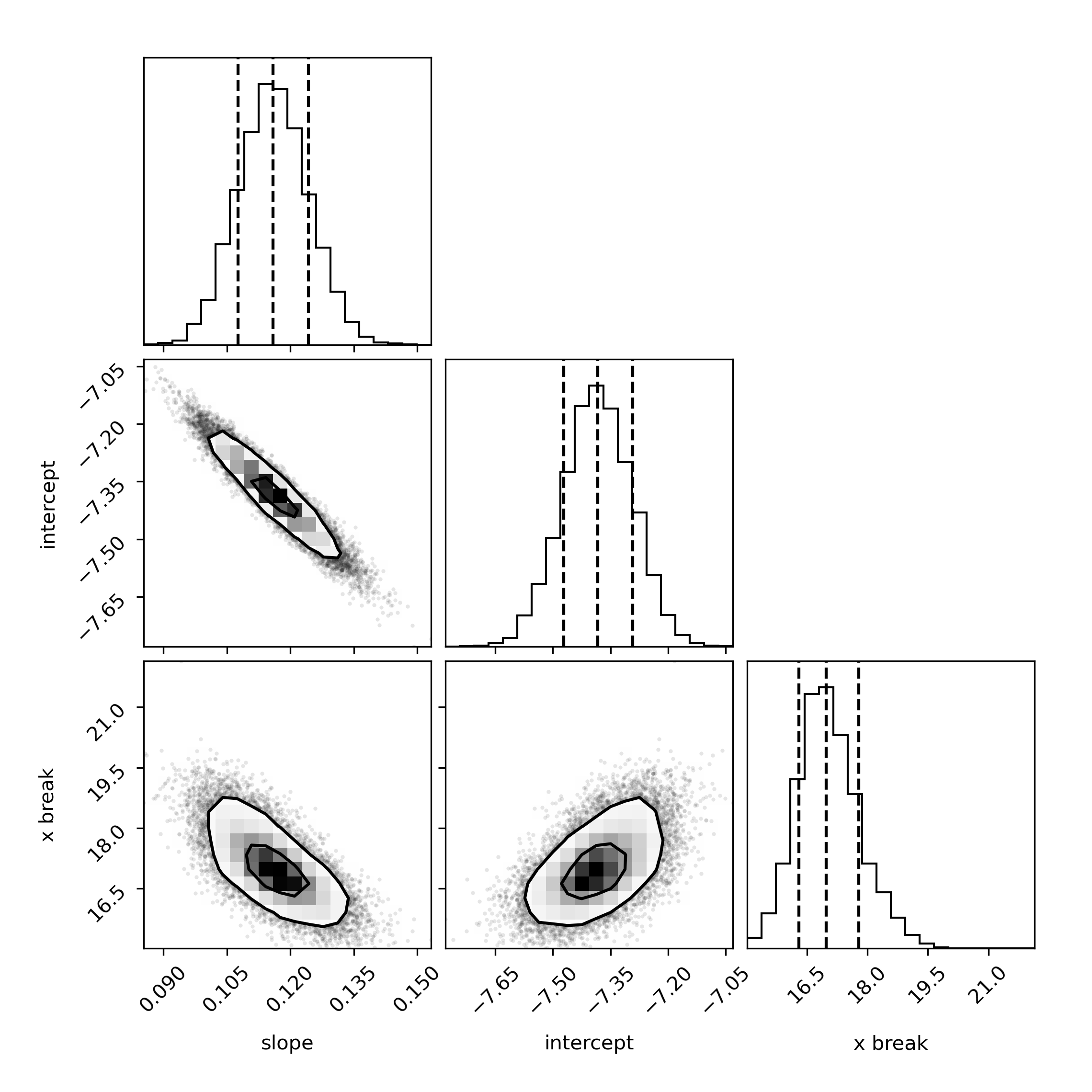}
    \caption{Corner plot of the MCMC fit to the log(MLR)-expansion velocity plot in Figure~\ref{fig:MLRvsvinf_lit}. The 16th, 50th, and 84th percentiles are marked with dashed lines, and the contour plots show 1$\sigma$ and 2$\sigma$ contours.}
    \label{fig:app:MLRvinfcorner}
\end{figure}

\clearpage

\section{Description of the literature sample for comparison with NESS results} \label{sect:lit_sample}

We will compare our observations to a combined literature sample of relatively large previous studies of AGB stars that have calculated gas MLRs, which includes the results of \citet{loup_co_1993, schoier_models_2001, olofsson_mass_2002, gonzalez_delgado_thermal_2003, ramstedt_circumstellar_2009, de_beck_probing_2010}. This sample has a total of 616 observational data points across $\sim$350 sources, comparable to our 1013 data points across 493 sources so far.  
This section will give an overview of the chosen samples of these studies, and summarize their results.

\citet{loup_co_1993} collect a sample of 444 evolved stars with previous detections in either CO or HCN. They calculate MLRs for 284 AGB sources, using an empirical formula based on CO (1--0) intensity from \citet{knapp_mass_1985}, and taking into account the CO dissociation radii calculated by \citet{mamon_photodissociation_1988}. In cases where they have several observations of a source, they report a mean calculated MLR and measured expansion velocity, but without an indication of the spectral quality of the various observations. As such, there are some cases of large discrepancies between individual values for a single source. 
For oxygen-rich AGB stars they find MLRs ranging from $1 \times 10^{-7}$ to $5 \times 10^{-5}$~M$_\odot$\,yr$^{-1}$, and expansion velocities generally between 5--20~km\,s$^{-1}$, with very few sources above 20--25~km\,s$^{-1}$. 
For carbon-rich AGB stars they find MLRs ranging from $3 \times 10^{-7}$ to $5 \times 10^{-5}$~M$_\odot$\,yr$^{-1}$, and expansion velocities generally between 5--30~km\,s$^{-1}$, with a fairly continuous distribution up to 35~km\,s$^{-1}$.
They caution that their sample has "no sound statistical basis but merely reflects the personal biases of the various observers in the field." It is obviously biased towards stronger sources (which tend to have higher MLRs) as they require a CO or HCN detection. The sample is also biased towards peculiar sources that previous observers have been interested in, such as bipolar outflows, which will make the MLR calculations less reliable. Their sample is also strongly biased against galactic plane sources and towards sources in the northern sky, though they note their inclusion of the \citet{nyman_survey_1992} SEST study helps mitigate this.

\citet{schoier_models_2001} observe a sample of 68 carbon-rich AGB stars, consisting of all sources with CO detections from an earlier study by \citet{olofsson_study_1993} which targeted the brightest (K < 2 mag) carbon stars in the sky. \citet{schoier_models_2001} note that their sample includes all sources with distances up to 500~pc, and is probably only missing about a third of sources out to the maximum distance of $\sim$1~kpc.
They use observations of CO (1--0), (2--1), and (3--2), and radiative transfer modelling to determine MLRs. 61 sources are well fit with a 1D model, and of the remaining 7 sources: 5 show detached shells, and 2 are not spherically symmetric. From the well-fit sources they derive MLRs between $5 \times 10^{-9}$ and $2 \times 10^{-5}$~M$_\odot$\,yr$^{-1}$, with a large fraction of sources around $3 \times 10^{-7}$~M$_\odot$\,yr$^{-1}$ and very few below $5 \times 10^{-8}$~M$_\odot$\,yr$^{-1}$. They say the lack of MLRs below $5 \times 10^{-8}$~M$_\odot$\,yr$^{-1}$ seems to be real, and probably indicates a lower limit to what is required to drive a dusty wind. 
They find that in general at high MLR the most important feature in determining the MLR is the temperature structure, while a wider range of parameters are important at low MLR. This makes sense as the CO emission tends to be saturated for high MLRs, making it harder to derive good model results from a few CO observations. 
They also find the MLR to be well correlated with the measured expansion velocity, and estimate that the studied types of carbon stars return $\sim$0.05~M$_\odot$\,yr$^{-1}$ of gas to the galaxy, while more extreme carbon stars (with MLRs above $2 \times 10^{-5}$~M$_\odot$\,yr$^{-1}$) may provide an order of magnitude more.

\citet{olofsson_mass_2002} present a sample of 69 oxygen-rich AGB stars, which are either semi-regular or irregular variables. These are the sources with CO detections from an earlier sample by \citet{kerschbaum_oxygen-rich_1999}, chosen based on IRAS colors indicating a dusty AGB envelope. They derive distances by assuming a luminosity of 4000~L$_\odot$. 
They use observations of CO (1--0), (2--1), (3--2), and (4--3), and 1D radiative transfer modelling to derive MLRs ranging from $2 \times 10^{-8}$ to $8 \times 10^{-7}$~M$_\odot$\,yr$^{-1}$, and find expansion velocities between 2.2 and 14.4~km\,s$^{-1}$. 30\% of their sources have expansion velocities below 5~km\,s$^{-1}$, so this sample seems to be biased towards slower winds compared to M stars in the NESS sample. 5 sources show expansion velocities below 3~km\,s$^{-1}$, which corresponds to the escape velocity at 100 R$_\star$, far beyond normally accepted acceleration zone which only extends to $\sim$20~R$_\star$. They speculate that this may be due to low radiative acceleration efficiency, leading to gas moving at a constant velocity from a few R$_\star$ and eventually escaping, yielding both low MLRs and low expansion velocities. 
In general they find a good correlation between MLR and expansion velocity.
They also compare their results with the \citet{schoier_models_2001} results described above, which were derived using the same methods, finding that the median MLRs are very similar. However \citet{olofsson_mass_2002} see a sharp cutoff around MLRs of $10^{-6}$~M$_\odot$\,yr$^{-1}$, which seems to be the maximum for these types of stars. They note, however, that their sample is biased by the IRAS colors selecting for dusty stars, and they of course do not include Mira variables which tend to have higher MLRs. Their range of MLRs also doesn't extend to values as low as those found for the carbon-rich sample.

\citet{gonzalez_delgado_thermal_2003} have a sample of 71 oxygen-rich AGB stars, which have all been detected in CO emission by \citet{kerschbaum_oxygen-rich_1999} and \citet{olofsson_mass_2002}. Using observations of several low-J transitions of SiO, for which they find a detection rate of $\sim$60\%, and radiative transfer modelling they derive MLRs and SiO radial abundance distributions for 44 sources. Additionally, they model CO (1--0), (2--1), (3--2), and (4--3) emission from the 12 Mira variables in their sample to derive MLRs.  
For these 12 sources they find a very high median MLR of $1.3 \times 10^{-5}$~M$_\odot$\,yr$^{-1}$, and only two sources have low MLRs around $10^{-7}$~M$_\odot$\,yr$^{-1}$. The median expansion velocity for the Miras is 15.3~km\,s$^{-1}$, also significantly higher than the 7~km\,s$^{-1}$ found by \citet{olofsson_mass_2002} for their sample of irregular and semiregular variables. Again, the two low MLR Miras are the only ones with expansion velocities below 10~km\,s$^{-1}$. 
Overall they find MLRs ranging from $2 \times 10^{-8}$ to $4 \times 10^{-5}$~M$_\odot$\,yr$^{-1}$, with a median value of $4 \times 10^{-7}$~M$_\odot$\,yr$^{-1}$. For expansion velocities they find a range from 2.3 to 19.3~km\,s$^{-1}$, with a median value of 7.5~km\,s$^{-1}$. 
They find that the SiO lines are generally narrower than CO, but have wide line wings meaning the measured expansion velocities from both SiO and CO lines are similar. 

\citet{ramstedt_circumstellar_2009} have a sample of 40 S-type AGB stars with previous CO detections, largely taken from the sample of \citet{jorissen_circumstellar_1998} of IRAS PSC S-type sources with good quality IRAS fluxes. They note their sample is likely biased towards higher MLRs, but they believe it to be representative of mass-losing S-type stars and complete to a distance of 600~pc (while the largest distance in their sample is 1210~pc). 
Using observations of CO (1--0) and (2--1), as well as one or more low-J transitions of SiO (which they detect in 26 sources), and 1D radiative transfer modelling they derive MLRs and CO and SiO radial abundance distributions. They find median MLRs of $4.5 \times 10^{-7}$~M$_\odot$\,yr$^{-1}$ and $1.75 \times 10^{-7}$~M$_\odot$\,yr$^{-1}$ for their Mira and SRV samples respectively. These numbers are comparable to the median value of $3 \times 10^{-7}$~M$_\odot$\,yr$^{-1}$ found for carbon-rich stars by \citet{schoier_models_2001} and for oxygen-rich stars by \citet{olofsson_mass_2002} and \citet{gonzalez_delgado_thermal_2003}. 
They find a median expansion velocity of 8~km\,s$^{-1}$, similar to the results for oxygen-rich sources, and slightly lower than the 11~km\,s$^{-1}$ found for carbon-rich sources in previous studies. 

\citet{de_beck_probing_2010} have a sample of 69 sources, including mostly AGB stars but also some RSGs, hypergiants, post-AGB stars, and YSOs. The data for this sample, consisting of $^{12}$CO and $^{13}$CO transitions up to J=6-5, has been assembled over many years. 
They use several CO transitions and 1D radiative transfer modelling to derive analytical expressions to estimate MLRs. They use this procedure to determine the MLRs for 50 of the evolved stars in the sample to which they could fit a soft parabola profile, of which 39 are AGB stars. They find AGB MLRs ranging from $4 \times 10^{-8}$ to $6 \times 10^{-5}$~M$_\odot$\,yr$^{-1}$, with a median value of $4.1 \times 10^{-6}$~M$_\odot$\,yr$^{-1}$. This is a higher median value than the $\sim$$3 \times 10^{-7}$~M$_\odot$\,yr$^{-1}$ found by the previously mentioned studies, indicating a significant bias towards bright sources in this sample. 
They find a correlation between MLR and pulsation period for periods below $\sim$850 days, representing Mira and semiregular AGB pulsators, as well as short-period OH/IR stars. 
For 29 stars in their sample they have both $^{12}$CO and $^{13}$CO observations, and hence are able to estimate $^{12}$CO/$^{13}$CO ratios by dividing their respective integrated intensities (with a correction factor for differences in line strength). They find $^{12}$CO/$^{13}$CO ratios ranging from 3.6 to 30.7 for their subsample of AGB stars (their Table 8), and note that these estimates are actually lower limits as the $^{12}$CO lines are often optically thick. 

Many other studies of samples of AGB stars also draw from the aforementioned surveys for their source selection \citep[e.g.,][]{teyssier_success_2011, massalkhi_abundance_2020, ramstedt_deathstar_2020}.
Overall this literature sample consists largely of sources with previous CO detections, so it is biased towards the brighter AGB stars with relatively high MLRs. The NESS sample thus complements these literature samples (Figure \ref{fig:dist-DPR}). Some of the samples of carbon-rich or S-type stars are said to be complete out to a few hundred pc, but there has been little attempt to form a complete sample of the much more abundant oxygen-rich stars, and this significantly hinders our ability to draw firm physical conclusions about them.

\clearpage

\section{Columns available in online table of results}

\begin{center}
\begin{table*}[ht]
\caption{Description of columns available in the online table}
\label{tab:analysis_results} 
\centering 
\begin{tabular}{lll}
\hline\hline
Column & Description & Unit\\
\hline
IRASPSC & IRAS PSC identifier & \\
SIMBAD\_ID & SIMBAD identifier & \\
Chem\_type & Chemical type based on mid-IR spectra [`O' or `C']; see Section 2.4 & \\
Tier & Grouping from Scicluna et al. (2022) based on location in distance-DPR space & \\
& (`very low', `low', `intermediate', `high', or `extreme') & \\
Peak\_CO(2-1)$^{a}$ & Peak intensity of the CO(2-1) line & K\\
Peak\_CO(2-1)\_error & Uncertainty in the peak intensity of the CO(2-1) line & K\\
v\_inf\_CO(2-1) & Expansion velocity of the shell derived from the CO(2-1) line (uncertainty 1 & km\,s$^{-1}$\\
& km/s) & \\
Int\_CO(2-1) & Integrated intensity of the CO(2-1) line & K km\,s$^{-1}$\\
Int\_CO(2-1)\_error & Uncertainty in the integrated intensity of the CO(2-1) line & K km\,s$^{-1}$\\
nchan\_CO(2-1) & Number of velocity channels in the CO(2-1) spectral band & \\
CO(2-1)\_rms & RMS noise in the CO(2-1) line & K\\
Int\_13CO(2-1) & Integrated intensity of the 13CO(2-1) line & K km\,s$^{-1}$\\
Int\_13CO(2-1)\_error & Uncertainty in the integrated intensity of the 13CO(2-1) line & K km\,s$^{-1}$\\
Peak\_CO(3-2) & Peak intensity of the CO(3-2) line & K\\
Peak\_CO(3-2)\_error & Uncertainty in the peak intensity of the CO(3-2) line & K\\
v\_inf\_CO(3-2) & Expansion velocity of the shell derived from the CO(3-2) line (uncertainty 1 & km\,s$^{-1}$\\
& km/s) & \\
Int\_CO(3-2) & Integrated intensity of the CO(3-2) line & K km\,s$^{-1}$\\
Int\_CO(3-2)\_error & Uncertainty in the integrated intensity of the CO(3-2) line & K km\,s$^{-1}$\\
nchan\_CO(3-2) & Number of velocity channels in the CO(3-2) spectral band & \\
CO(3-2)\_rms & RMS noise in the CO(3-2) line & K\\
Int\_13CO(3-2) & Integrated intensity of the 13CO(3-2) line & K km\,s$^{-1}$\\
Int\_13CO(3-2)\_error & Uncertainty in the integrated intensity of the 13CO(3-2) line & K km\,s$^{-1}$\\
MLR\_CO(2-1) & Empirical mass-loss rate derived from the CO(2-1) line using the Ramstedt et al. & M$_\odot$\,yr$^{-1}$\\
& (2008) formula & \\
MLR\_CO(2-1)\_error & Uncertainty in the mass-loss rate derived from the CO(2-1) line & M$_\odot$\,yr$^{-1}$\\
CO(2-1)\_upperlimit & Flag designating whether the CO(2-1) line is detected (0) or is below 3 times & \\
& the RMS (1) & \\
MLR\_CO(3-2) & Empirical mass-loss rate derived from the CO(3-2) line using the Ramstedt et al. & M$_\odot$\,yr$^{-1}$\\
& (2008) formula & \\
MLR\_CO(3-2)\_error & Uncertainty in the mass-loss rate derived from the CO(3-2) line & M$_\odot$\,yr$^{-1}$\\
CO(3-2)\_upperlimit & Flag denoting whether the CO(3-2) line is detected (0) or is below 3 times the & \\
& RMS (1) & \\
IsoRatio(2-1) & Isotopic ratio 12CO/13CO derived from the (2-1) lines & \\
IsoRatio(3-2) & Isotopic ratio 12CO/13CO derived from the (3-2) lines & \\
DPR & Dust-production rate from SED fit & M$_\odot$\,yr$^{-1}$\\
DPR\_error & Uncertainty in the dust-production rate & M$_\odot$\,yr$^{-1}$\\
GasDust\_CO(2-1) & Gas-to-dust ratio derived from the CO(2-1) line & \\
GasDust\_CO(2-1)\_error & Uncertainty in the gas-to-dust ratio derived from the CO(2-1) line & \\
GasDust\_CO(3-2) & Gas-to-dust ratio derived from the CO(3-2) line & \\
GasDust\_CO(3-2)\_error & Uncertainty in the gas-to-dust ratio derived from the CO(3-2) line & \\
SoftParabola\_CO(2-1) & Flag denoting whether the CO(2-1) line shape is a soft parabola (1) or not (0) & \\
SoftParabola\_CO(3-2) & Flag denoting whether the CO(3-2) line shape is a soft parabola (1) or not (0) & \\
OptThin\_12CO(2-1) & Flag denoting whether the CO(2-1) line is optically thin (1) or not (0) & \\
OptThin\_12CO(3-2) & Flag denoting whether the CO(3-2) line is optically thin (1) or not (0) & \\
First\_det & Flag denoting whether the NESS observation is the first CO detection for the & \\
& source (1) or not (0) & \\
\hline
\multicolumn{3}{l}{Notes: $^{a}$ All peak intensities, expansion velocities, and the related uncertainties in the table are estimated from soft-parabola fits.}\\\multicolumn{3}{l}{Integrated intensities are derived from the spectrum, not from the models.}\\
\hline
\end{tabular} 
\end{table*}
\end{center}

%%%%%%%%%%%%%%%%%%%%%%%%%%%%%%%%%%%%%%%%%%%%%%%%%%

\end{document}